\documentclass[structabstract]{aa}
\usepackage[modulo,switch]{lineno}
\usepackage{graphicx}
\usepackage{natbib}
\usepackage{longtable}
\usepackage{longtable,lscape}
\usepackage{txfonts}
\begin{document}
\linenumbers
\newcommand{\meandnu} {\langle\Delta\nu\rangle}
   \title{Asteroseismic inferences on red giants in open clusters NGC~6791, NGC~6819 and NGC~6811 using \textit{Kepler}}

   \author{S. Hekker\inst{1,2} \and S. Basu\inst{3}  \and D. Stello\inst{4} \and T. Kallinger\inst{5} \and F. Grundahl\inst{6} \and S. Mathur\inst{7} \and R.A. Garc\'ia\inst{8} \and B. Mosser\inst{9} \and D. Huber\inst{4} \and T.R. Bedding\inst{4} \and  R. Szab\'o\inst{10} \and J. De Ridder\inst{11} \and W.J. Chaplin\inst{2} \and Y. Elsworth\inst{2} \and S.J. Hale\inst{2} \and J. Christensen-Dalsgaard\inst{6} \and R.L. Gilliland\inst{12} \and M. Still\inst{13} \and S. McCauliff\inst{14} \and E.V. Quintana\inst{15}}

\offprints{S. Hekker, \\
                    email: S.Hekker@uva.nl}
   \institute{Astronomical Institute "Anton Pannekoek", University of Amsterdam, Science Park 904, 1098 XH Amsterdam, the Netherlands
   \and School of Physics and Astronomy, University of Birmingham, Edgbaston, Birmingham B15 2TT, United Kingdom
   \and Department of Astronomy, Yale University, P.O. Box 208101, New Haven CT 06520-8101, USA
   \and Sydney Institute for Astronomy (SIfA), School of Physics, University of Sydney, NSW 2006, Australia
   \and Department of Physics and Astronomy, University of British Colombia, 6224 Agricultural Road, Vancouver, BC V6T 1Z1, Canada
   \and Department of Physics and Astronomy, Building 1520, Aarhus University, 8000 Aarhus C, Denmark
   \and High Altitude Observatory, NCAR, P.O. Box 3000, Boulder, CO 80307, USA
   \and Laboratoire AIM, CEA/DSM-CNRS, Universit\'{e} Paris 7 Diderot, IRFU/SAp, Centre de Saclay, 91191, GIf-sur-Yvette, France
   \and LESIA, UMR8109, Universit\'e Pierre et Marie Curie, Universit\'e Denis Diderot, Observatoire de Paris, 92195 Meudon Cedex, France
   \and Konkoly Observatory of the Hungarian Academy of Sciences, Konkoly Thege Mikl\'os \'ut 15-17, H-1121 Budapest, Hungary
   \and Instituut voor Sterrenkunde, K.U. Leuven, Celestijnenlaan 200D, 3001 Leuven, Belgium
   \and Space Telescope Science Institute, 3700 San Martin Drive, Baltimore, MD 21218, USA
   \and Bay Area Environmental Research Institute / Nasa Ames Research Center, Moffett Field, CA 94035, USA
   \and Orbital Sciences Corporation / Nasa Ames Research Center, Moffett Field, CA 94035, USA
   \and SETI Institute / Nasa Ames Research Center, Moffett Field, CA 94035, USA\\
         }

   \date{Received ; accepted}


  \abstract
   {Four open clusters are present in the \textit{Kepler} field of view and timeseries of nearly a year in length are now available. These timeseries allow us to derive  asteroseismic global oscillation parameters of red-giant stars in the three open clusters NGC~6791, NGC~6819 and NGC~6811. From these parameters and effective temperatures, we derive mass, radii and luminosities for the clusters as well as field red giants. }
   {We study the influence of evolution and metallicity on the observed red-giant populations. }
   {The global oscillation parameters are derived using different published methods and the effective temperatures are derived from 2MASS colours. The observational results are compared with \textit{BaSTI} evolution models.}
   {We find that the mass has significant influence on the asteroseismic quantities $\Delta \nu$ vs. $\nu_{\rm max}$ relation, while the influence of metallicity is negligible, under the assumption that the metallicity does not affect the excitation / damping of the oscillations. The positions of the stars in the H-R diagram depend on both mass and metallicity. Furthermore, the stellar masses derived for the field stars are bracketed by those of the cluster stars.}
   {Both the mass and metallicity contribute to the observed difference in locations in the H-R diagram of the old metal-rich cluster NGC~6791 and the middle-aged solar-metallicity cluster NGC~6819. 
For the young cluster NGC~6811, the explanation of the position of the stars in the H-R diagram challenges the assumption of solar metallicity, and this open cluster might have significantly lower metallicity [Fe/H]  in the range $-$0.3 to $-$0.7 dex. Also, nearly all the observed field stars seem to be older than NGC~6811 and younger than NGC~6791.}

   \titlerunning{Asteroseismic properties of red giants in open clusters NGC~6791, NGC~6819 and NGC~6811}
   \authorrunning{Hekker et al.}
   \maketitle
%

\section{Introduction}
Asteroseismology is a powerful tool to obtain information about the internal structure of stars. For stars exhibiting solar-like oscillations, i.e., stars with a turbulent outer layer, global oscillation parameters can be used to derive the mass and radius of a star. This provides preliminary information on the evolutionary state of a star and for an ensemble of stars this can be used to investigate the population structure \citep[e.g.][]{miglio2009}.
Using asteroseismology to its full extent and deriving the internal structure of stars in detail from the observed oscillations requires accurate knowledge of atmospheric stellar parameters, such as effective temperature, surface gravity, luminosity and metallicity. Many projects are striving to determine these atmospheric stellar parameters with the necessary accuracy \citep[e.g.][]{uytterhoeven2010, molenda2010}, which is an immense task for the large number of relatively faint stars for which we currently have asteroseismic results thanks to the CoRoT \citep{baglin2006} and \textit{Kepler} \citep{borucki2009} satellites. Note that both CoRoT and \textit{Kepler} consortia provide large databases with atmospheric stellar parameters of all observed stars, i.e. EXODAT \citep{meunier2007} and the \textit{Kepler} Input Catalogue \citep[KIC,][]{brown2011}. However, these data are primarily collected for exo-planet target selection and the accuracy and precision of these data are often not sufficient for asteroseismic purposes. 

There are limitations to the accuracy with which atmospheric stellar parameters can be obtained for single field stars. For stars in a binary or in a cluster additional constraints are available from the binary or cluster nature. In the work presented here we focus on clusters and will take advantage of having both field and cluster red giants in the \textit{Kepler} field of view. All stars in a certain cluster are formed at the same time from the same cloud of gas and dust and therefore we assume that all stars in the cluster have the same age, distance and metallicity. This knowledge is used in the asteroseismic investigation of the clusters. Several authors have already endeavoured to perform asteroseismic investigations of cluster stars. Indeed, evidence for solar-like oscillations in K giants has been reported in the open clusters M67 \citep{stello2007} and with the first \textit{Kepler} data in NGC~6819 \citep{stello2010}, and the globular clusters 47 Tuc \citep{edmonds1996} and NGC~6397 \citep{stello2009}. However, the asteroseismic data obtained for cluster stars prior to \textit{Kepler} could only reveal the presence of the oscillations, but were not precise enough to obtain reliable global oscillation parameters. Results for field red giants observed using CoRoT or \textit{Kepler} data have already been presented by \citet{deridder2009, hekker2009,kallinger2010a,mosser2010,bedding2010,hekker2010,huber2010,kallinger2010}.

We concentrate here on new data from the NASA \textit{Kepler} mission. This NASA mission was launched successfully in March 2009 and is taking data of unprecedented quality in a large field of 105-square degrees in the direction of Cygnus and Lyra. In this field there are four open clusters present. For three of them, red-giant stars with magnitudes $V_{\rm RG}$ in the range observable with \textit{Kepler} are present: NGC~6811 (9.5~$<$~V$_{\rm RG}$~$<$~12.0), NGC~6819  (13.5~$<$~V$_{\rm RG}$~$<$~14.3) and NGC~6791 (13.7~$<$~V$_{\rm RG}$~$<$17.5). NGC~6811 has an age of 0.7 $\pm$ 0.1 Gyr \citep{glushkova1999} and contains only a handful of red giants. The older clusters NGC~6819 and NGC~6791 with ages of about 2.5 and 10 Gyr, respectively, have a considerable population of red-giant stars. The fourth and youngest cluster NGC~6866 has an age of about 0.56 Gyr \citep{frolov2010} and no red giants have been observed in this cluster.

\subsection{NGC~6791}
NGC~6791 is one of the oldest, most massive and most metal-rich open clusters known \citep{origlia2006,carretta2007,anthony2007}, and contains a population of hot blue stars \citep{liebert1994,landsman1998} and white dwarfs extending to the end of the cooling sequence \citep{bedin2005,bedin2008,kalirai2007}. For these reasons NGC~6791 has been studied extensively. Nevertheless, little agreement  concerning its basic parameters has been reached. Normally, a colour-magnitude diagram is used to derive the age of the cluster. The non-negligible reddening increases the uncertainty of the age for NGC~6791. Therefore, other probes have been used, such as eclipsing binary systems \citep[see e.g.,][]{grundahl2008,brogaard2011} and the white dwarf cooling sequence \citep{bedin2005,bedin2008,kalirai2007}.

The ages proposed for NGC~6791 range from 7 to 12 Gyr \citep[see e.g.,][]{basu2011,grundahl2008}, which is longer than the dynamical relaxation time, i.e., the time in which individual stars exchange energies and thier velocity distribution approaches a Maxwellian equilibrium. Thus, NGC~6791 is dynamically relaxed \citep{durgapal2001}. In addition, there are four independent studies available to determine the metallicity of this cluster. They found [Fe/H] = +0.39~$\pm$~0.05 \citep[][high-resolution spectroscopy]{carraro2006}, [Fe/H] = +0.35~$\pm$~0.02 \citep[][high-resolution spectroscopy]{origlia2006}, [Fe/H] = +0.45~$\pm$~0.04 \citep[][multi-colour photometry]{anthony2007} and [Fe/H] = +0.29~$\pm$~0.03 (random) $\pm$~0.07 (systematic) \citep[][spectroscopy]{brogaard2011}. As pointed out by \citet{carretta2007}, comparing these values is complicated as different subsets of stars are observed in each study. It seems likely that the differences are mainly caused by differences in the adopted reddening, i.e., either $E(B-V)$ = 0.09 \citep{stetson2003} or $E(B-V)$ = 0.15 \citep[an average of literature determinations,][]{carretta2007}. These reddening values are used to derive atmospheric stellar parameters from photometry. The resulting different atmospheric stellar parameters are used in the different spectroscopic metallicity studies. 

\begin{figure}
\begin{minipage}{\linewidth}
\centering
\includegraphics[width=\linewidth]{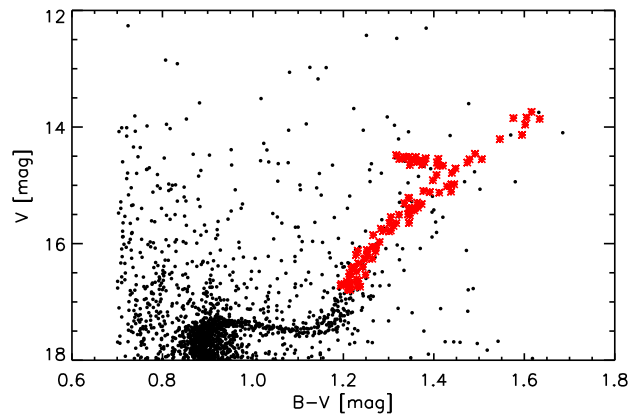}
\end{minipage}
\caption{Colour-magnitude diagram of NGC~6791 \citep{stetson2003}. Target stars used in the present study are indicated with red asterisks.}
\label{CMD6791}
\end{figure}

\subsection{NGC~6819}
NGC~6819 is a very rich open cluster with roughly solar metallicity of [Fe/H] = +0.09 $\pm$ 0.03 \citep{bragaglia2001}, an age of about 2.5 Gyr \citep{kalirai2001,kalirai2004} and reddening $E(B-V)$ = 0.15. There is reasonable agreement on the metallicity, reddening and age of this cluster and therefore NGC~6819 has been used to study other phenomena, such as the initial-final mass relation using the population of white dwarfs present in this cluster \citep[e.g.][]{kalirai2008}. There is also clear evidence for mass segregation in NGC~6819, i.e., the  giants and upper main-sequence stars are concentrated in the inner regions, whereas the lower main-sequence stars distribute almost uniformly throughout the cluster. This results from the fact that the age of NGC~6819 is about 10 times larger than its dynamical relaxation time \citep{kang2002}. 

\subsection{NGC~6811}
NGC~6811 is a young, sparse, not particularly well studied cluster. Studies on this cluster tended to focus on membership or variability of stars and no direct metallicity studies are available. Solar metallicity has been used as an initial guess. \newline

\subsection{Current investigation}
In the present study we compare asteroseismic global parameters $\nu_{\rm max}$ (the frequency of maximum oscillation power) and $\Delta \nu$ (the large frequency spacing between modes of the same degree and consecutive order) of solar-like oscillations in red-giant stars in the three open clusters NGC~6791, NGC~6819 and NGC~6811, and of field giants, all observed with \textit{Kepler}.  In this paper, we subsequently derive stellar parameters, such as mass and radius, from the asteroseismic parameters and investigate the influence of evolution and metallicity on the observed red-giant populations. The \textit{Kepler} data of the cluster stars are also being used to investigate mass loss along the red-giant branch (Miglio et al. in preparation), the ages of the clusters \citep{basu2011} and cluster membership \citep{stello2011}. The latter will be an extension of the study of NGC~6819 recently presented by \citet{stello2010}. 

\begin{table}
\begin{minipage}{8.4cm}
\caption{Adopted parameters, and data sources for the clusters in the present study.}
\label{cluster_param}
\centering
\begin{tabular}{lcccl}
\hline\hline
\tiny{cluster}    &  \tiny{[Fe/H]} &  \tiny{$E(B-V)$}   &   \tiny{$E(V-K)$}  & \tiny{V\_source}\\
\hline
\tiny{NGC~6791}  &  \tiny{$+$0.30$\pm$0.1} &  \tiny{0.16$\pm$0.02}\footnote{\citet{brogaard2011}}  &   \tiny{0.432}  & \tiny{\citet{stetson2003}}\\
\tiny{NGC~6819}  &  \tiny{0.0$\pm$0.1} &  \tiny{0.15}  &          \tiny{0.405} &  \tiny{\citet{hole2009}}\\
\tiny{NGC~6811}  &  \tiny{0.0$\pm$0.1} &  \tiny{0.16}  &          \tiny{0.435} &  \tiny{Webda}\footnote{http://www.univie.ac.at/webda/}\\
\hline
\end{tabular}
\end{minipage}
\end{table}

\begin{figure*}
\begin{minipage}{8.4 cm}
\centering
\includegraphics[width=8.4cm]{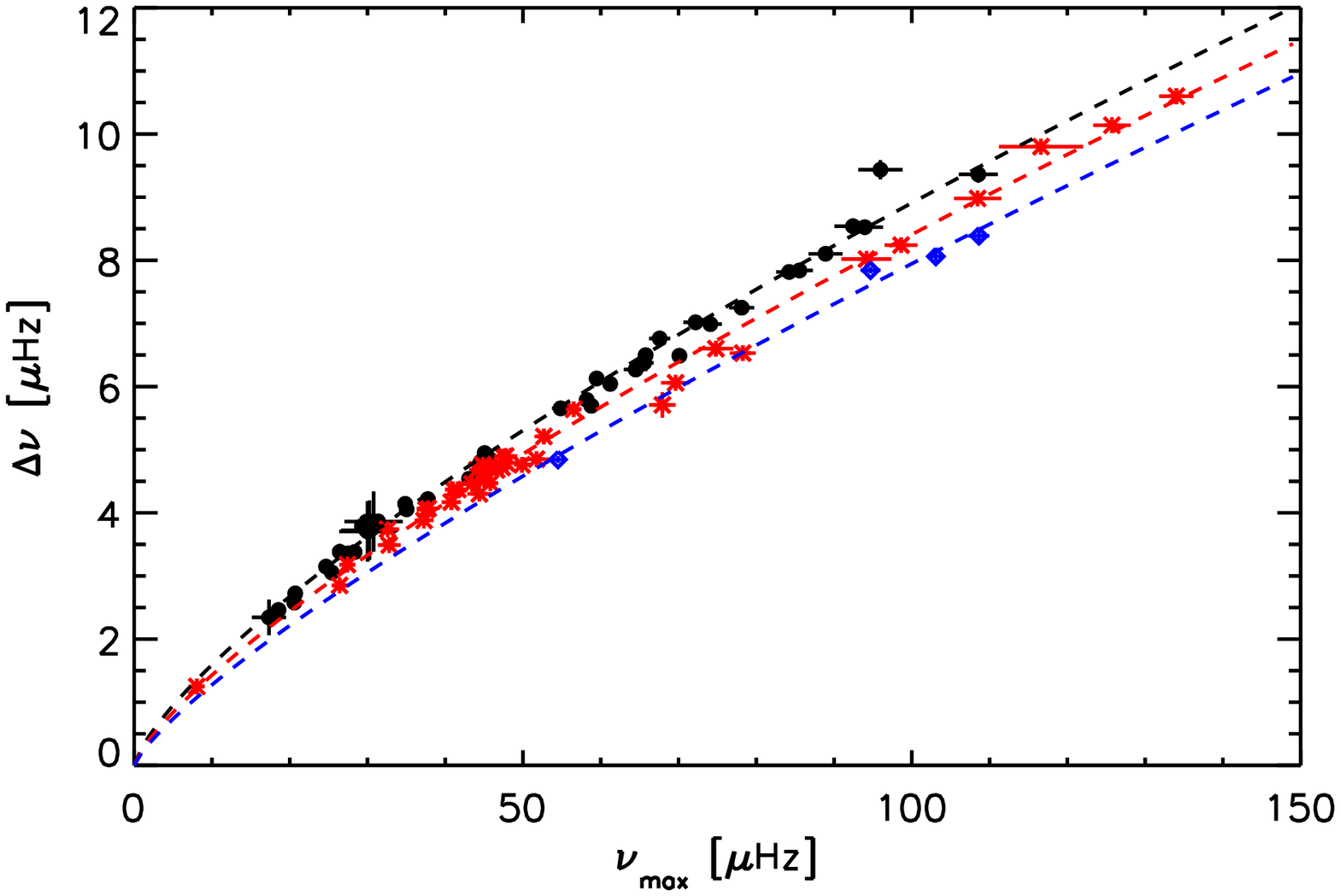}
\end{minipage}
\hfill
\begin{minipage}{8.4 cm}
\centering
\includegraphics[width=8.4cm]{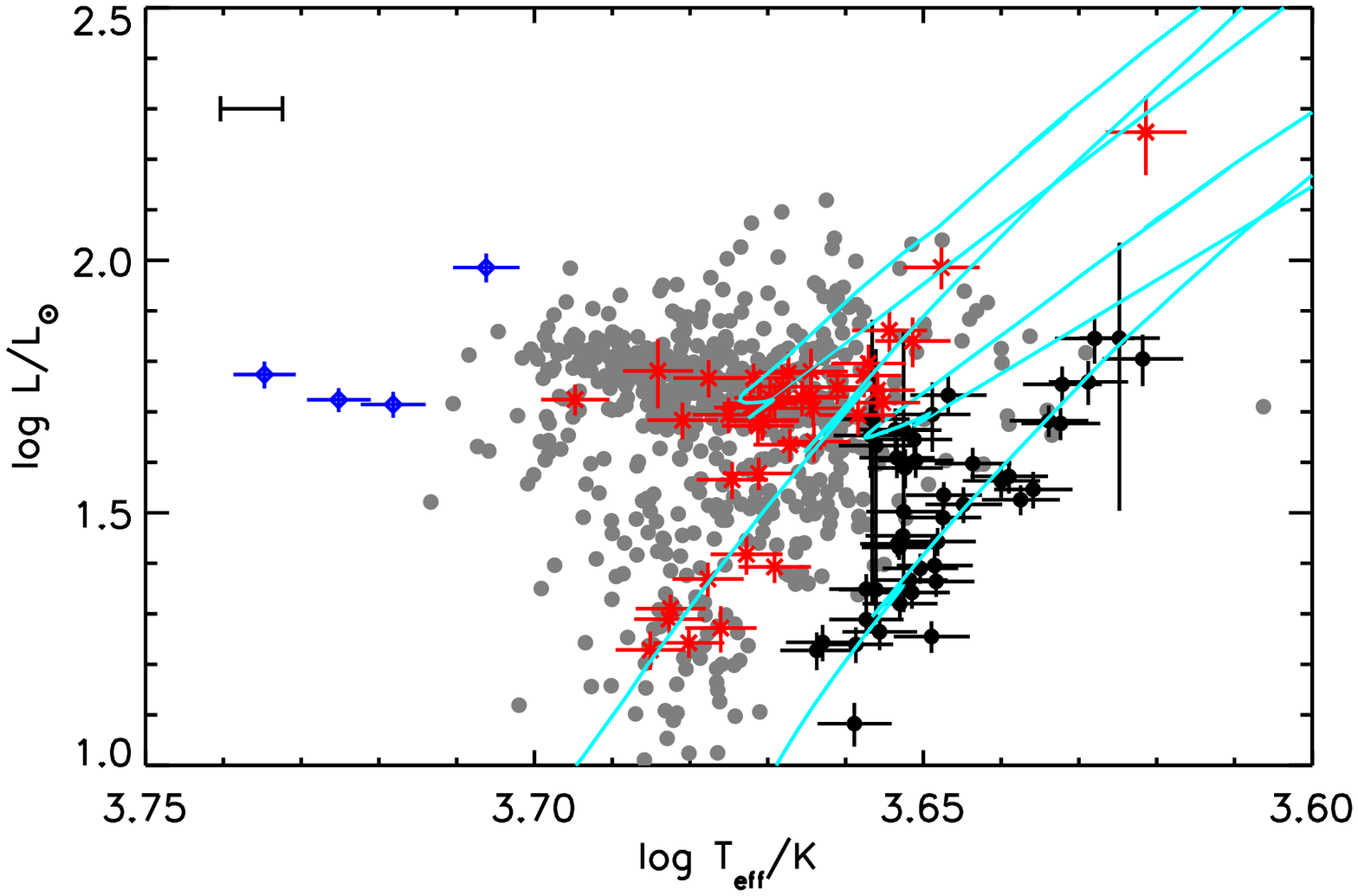}
\end{minipage}
\hfill
\begin{minipage}{8.4cm}
\centering
\includegraphics[width=8.4cm]{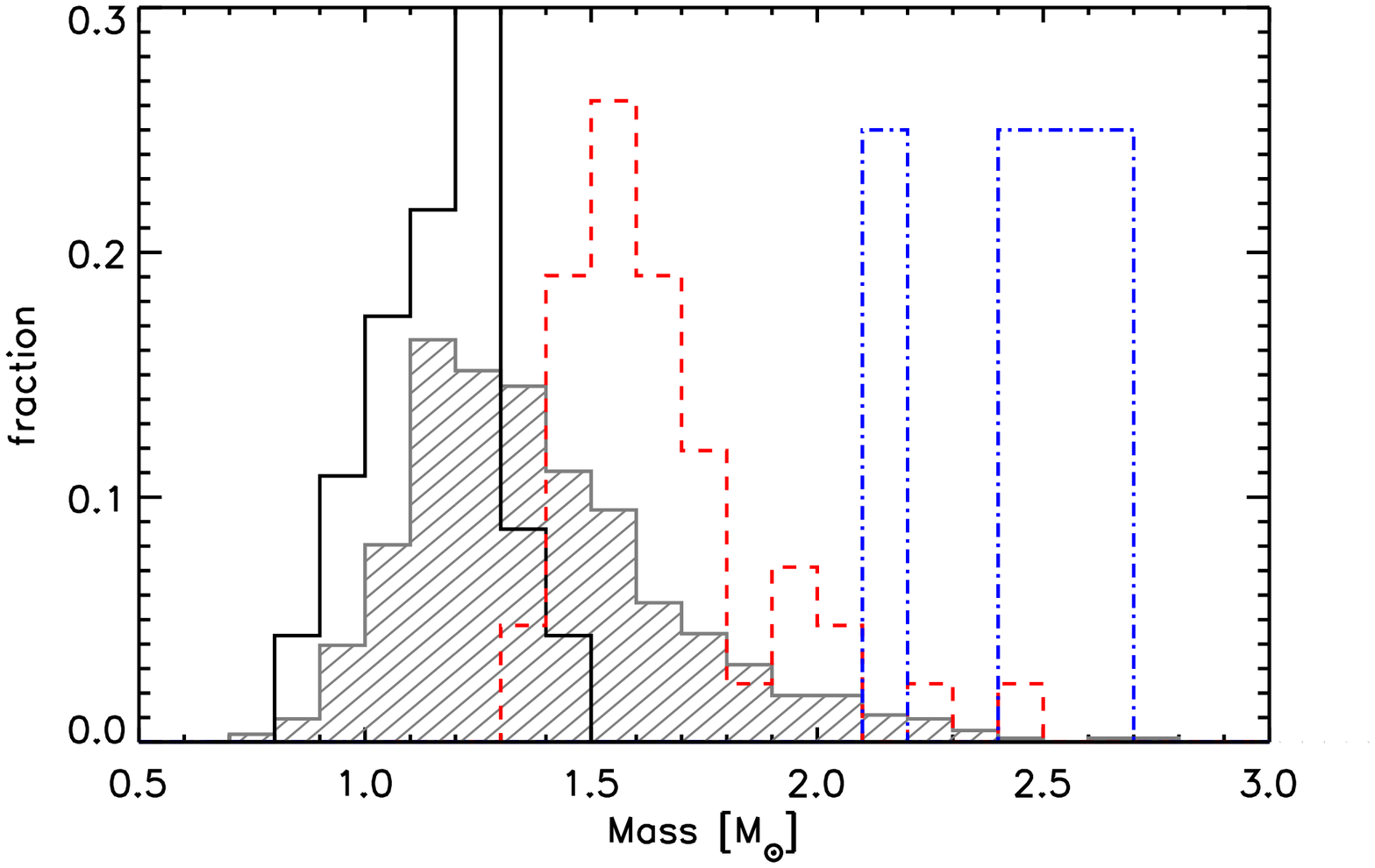}
\end{minipage}
\hfill
\begin{minipage}{8.4cm}
\centering
\includegraphics[width=8.4cm]{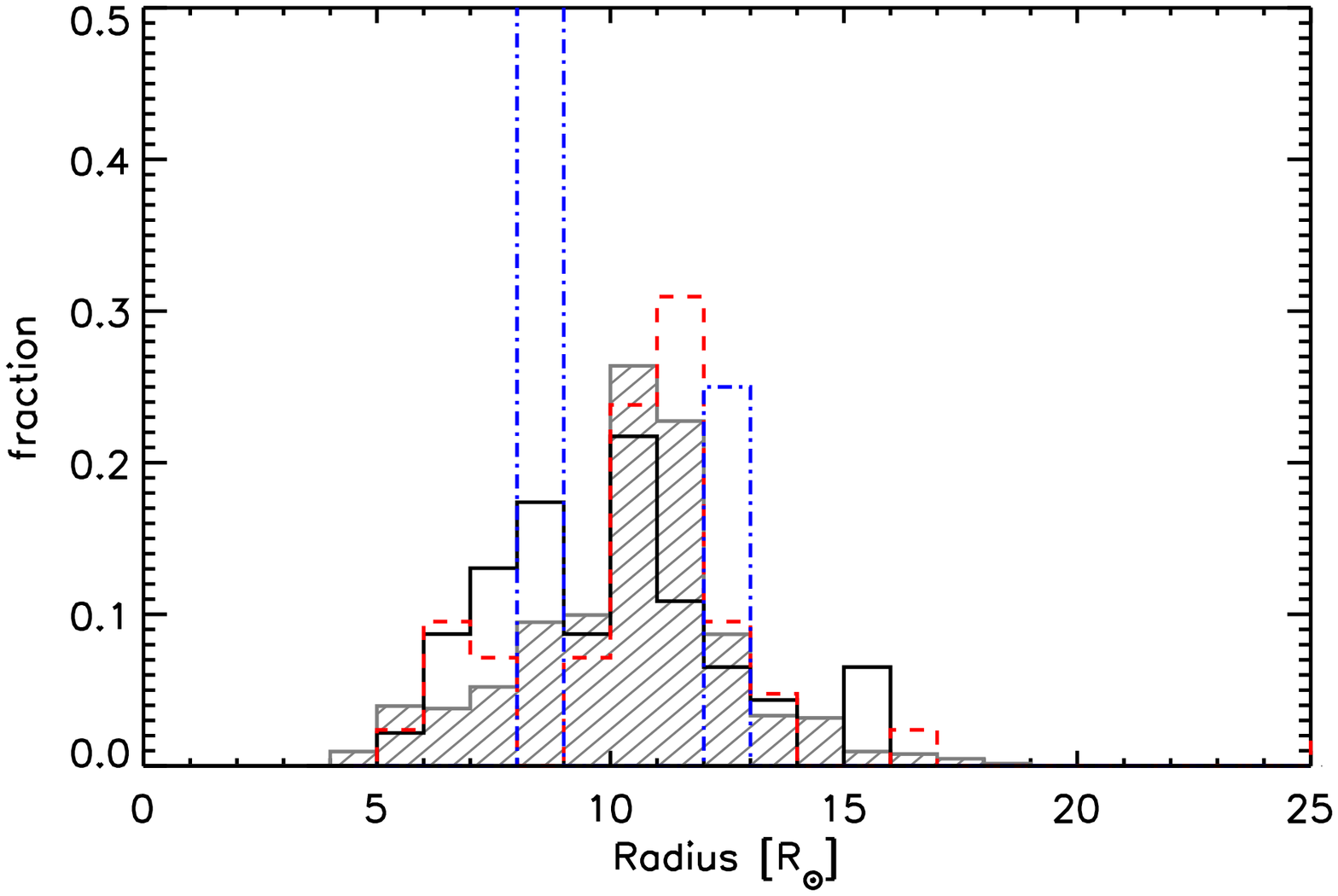}
\end{minipage}
\caption{Observed and derived parameters for the clusters NGC~6791 (black), NGC~6819 (red) and NGC~6811 (blue). The top panels show $\Delta \nu$ vs. $\nu_{\rm max}$ (left) and the derived luminosity vs $T_{\rm eff}$ (right). The uncertainties in the individual points in the H-R diagram are only the intrinsic uncertainties (50~K, see text), the possible offset due to systematic uncertainties in the effective temperatures (110~K) is indicated with the black horizontal bar in the top left corner. Isochrones for NGC~6791 and NGC~6819 are shown in cyan. The results for the field stars are shown in gray without uncertainties. The histograms of the asteroseismic mass (left) and radius (right) are shown in the bottom panels for the clusters NGC~6791 (black solid line), NGC~6819 (red dashed line), NGC~6811 (blue dotted-dashed line) and in gray hatched for the field stars.}
\label{resclusters}
\end{figure*}

\section{Observations}

\subsection{Cluster target selection}

To select the stars in NGC~6819 we used the radial velocity study by \citet{hole2009}.  It gives membership probabilities for all stars in the cluster vicinity down to $V=15.0$, which includes all stars targeted in this paper.  All stars with high membership probability ($P_{\mathrm{RV}}>80\%$) were chosen.  With this purely kinematic criterion we avoid any biases in our selection that could otherwise affect the distributions of the measured atmospheric stellar parameters. Due to the lack of a similarly complete kinematic membership study for
NGC~6791, we followed a slightly different selection procedure for this cluster. Based on the photometric study by \citet{stetson2003} we selected stars that were clearly photometric members, meaning they are located close to the cluster sequence including the red-giant branch and the red clump (see Fig.~\ref{CMD6791}). While this does not introduce significant biases in selecting particular populations of stars within the cluster, it does have the potential of leaving out members that are somewhat off the prime cluster sequence.
However, this selection effect is not likely to play a major role in the interpretation of the results in this paper. Finally, the stars selected in NGC~6811 were the only red giant 
candidate members from a preliminary radial velocity study of the cluster (Meibom, priv. comm.). 

We note here that for all clusters, non-members based on the asteroseismic studies by \citet{stello2010,stello2011}, have been removed from the present investigation.

\subsection{\textit{Kepler} data}
Timeseries data obtained during the first one-month (Q1) and subsequent three-month quarterly rolls (Q2, Q3, Q4) with the \textit{Kepler} satellite in long-cadence mode \citep[][29.4-minute near-equidistant sampling]{jenkins2010a} are used to obtain the asteroseismic parameters. The \textit{Kepler} data suffer from some instrumental effects. The most prominent effects are caused by 1) downlinking science data when the spacecraft changes its attitude, 2) safe-mode events during which the satellite warms up and the subsequent thermal relaxation affects the photometry, and 3) attitude adjustments necessary to compensate for drift of the satellite. The first two effects cause a gradual thermal drift that diminishes over time, while the latter effect causes jumps in the data. The \textit{Kepler} mission Science Operations Centre developed software to eliminate these effects \citep[e.g.][]{jenkins2010} in the Pre-search Data Conditioning (PDC) process. However, this software was not designed specifically for asteroseismic purposes and in some cases the signal from the oscillations and the related granulation were affected by the PDC corrections, which we took into account in the following way.
In these cases either the raw data were used, data were corrected in the same manner as the field stars \citep[see below and][]{garcia2011}, data were high-pass filtered, bad points were removed or a polynomial fit was removed, either for the complete timeseries or for individual segments separately. 

The timeseries of the field stars have been corrected with a different philosophy in which we try to preserve, as much as possible, the low-frequency trends that could have a stellar origin. These corrections are applied to the raw data and based on principles developed for GOLF/SoHO thermal and high voltage corrections \citep{garcia2005}. The corrections contain three steps and are described in more detail by \citet{garcia2011}. First, a multiplicative correction is performed for the ranges affected by thermal drift using a third order polynomial in the affected part and a second order polynomial in the adjacent part(s) of the timeseries. Then, jumps are detected from spurious differences in the mean power of adjacent segments of the light curve spanning one day. Additionally, we checked for the presence of jumps at known times of attitude adjustments. 

Point-to-point sigma clipping has been applied with a sigma clipping threshold chosen in combination with the other corrections applied to the data and depends on the sensitivity of the correction method. It is important to note that only a handful of data points (out of the $\sim$14000 points of the time series) are affected by the sigma-clipping.

Tests have been performed to investigate potential differences in the resulting power spectra of all stars with light curves corrected with the two different methods. Although there are differences in the individual frequency peaks, the $\Delta \nu$ and $\nu_{\rm max}$ are the same for data corrected with both methods. 

\section{Estimates of $T_{\rm eff}$ for the stars}

To estimate the effective temperatures for our targets we use the colour--temperature calibrations by \citet[][hereafter RM05]{ramirez2005}. All targets are sufficiently bright to have $JHK$ photometry from the 2MASS catalog \citep{skrutskie2006} allowing us to determine the temperatures based on the $(V-K)$ colour.
For each cluster we obtained the $V$ band photometry and the reddening from the sources given in Table~\ref{cluster_param}. 
Using the RM05 colour--temperature relations we find that for the giants an error of 0\fm02 in $(V-K)$ leads to an error of 15~K. An error of 
0.1~dex in [Fe/H] leads to an error in the temperature of $\sim$5~K. 

We adopt a random error of 0.01 for the $V$ magnitude for all stars, which we consider a safe upper limit for most stars (particularly in NGC~6791). The 2MASS catalog \citep{skrutskie2006} provides an error estimate for the $K$ magnitude of each star, for the ($V-K$) error we shall add these two in quadrature. For NGC~6791 the median of the ($V-K$) errors is 0.026, or 20~K. 

In addition to the random errors we also try to assign an estimate of the possible bias in the temperatures for the stars. This reflects our inability to properly account for the correct zeropoint of the temperature scale. The contributions to the bias come from: zeropoint errors in the photometry, reddening and the temperature calibration used. We consider the bias to have a `box-like' (= uniform) distribution and provide below the estimated half-length of the box.

For the $V$ and $K$ photometry we shall adopt a value of 0.02~mag ($\sim$15~K) for each filter. This is our estimate of how well our $V$ and $K$ magnitudes are transformed to the standard system (filter, CCD and stellar types are typically different from the instruments defining the standard system). See \citet{stetson2003} for a discussion.

The reddening estimate also carries a bias, which we estimate to be 0.02~mag in $E(B-V)$ -- leading to 0.054~mag in $E(V-K)$, or $\sim$40~K. We do not consider the $\pm0.02$ error reported for NGC~6791 in \citet{brogaard2011} to be a random error; this is because reddening estimates are often based on several different methods which may carry systematic inherent differences. Furthermore, for NGC~6791 the estimates of reddening have a long and varied history, essentially spanning values between 0.1 and 0.2 in $E(B-V)$ \citep[see][]{stetson2003}. Of the three clusters studied here, NGC~6791 is the best studied. Finally, we also include a possible bias due to the specific choice of colour-$T_{\rm eff}$ relation used here. RM05 gives a discussion of their calibration in comparison to other studies. Although it is difficult to give a single value based on their discussion we shall here adopt a value of 40~K. 

To arrive at the final bias estimate we thus have: 15~K (V) + 15~K (K) + 40~K (reddening) + 40~K (RM05) = 110~K. This is clearly dominant over the random errors which range between 15~K and 50~K for stars in NGC~6791. Thus we shall overall adopt random (star-to-star) errors which are smaller than 50~K, whereas our estimates above suggest that a bias up to $\sim$110~K in the zeropoint of the temperatures within a cluster can be a real possibility.\newline

For the field stars we use the effective temperature from the KIC \citep{brown2011} with a total uncertainty estimate of 150~K.

\begin{figure}
\begin{minipage}{\linewidth}
\centering
\includegraphics[width=\linewidth]{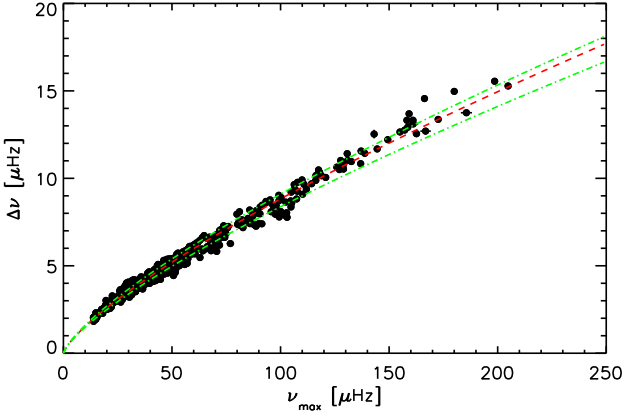}
\end{minipage}
\caption{$\Delta \nu$ vs. $\nu_{\rm max}$ for field red giants with the fit through all the data in red. The upper green dashed line is a for the low-mass sample and the lower green dashed line is a fit for the high mass stars (see text for an explanation of the samples).}
\label{resfield}
\end{figure}

\section{Results}
The timeseries, prepared as described above, were analysed for global oscillation parameters, i.e. the frequency of maximum oscillation power ($\nu_{\rm max}$) and frequency separation between modes of the same degree but consecutive orders ($\Delta \nu$), using five different methods, namely SYD \citep{huber2009}, COR \citep{mosser2009}, CAN \citep{kallinger2010a}, A2Z \citep{mathur2010} and OCT \citep{hekker2010b}. For a comparison of these methods see \citet{hekker2010}. We first looked at all five sets of results and computed for each star the (unweighted) mean and variance of both $\Delta \nu$ and $\nu_{\rm max}$. Only stars for which at least four methods produced consistent results for both parameters were taken into account. We use the unweighted means because different methods give very different estimates of the formal errors in each quantity. However, the results obtained from the different methods show that, the results themselves are not widely scattered. Hence we think it is justified to continue with the following procedure: for each cluster and for the field stars separately, we computed the total squared deviation from the means, i.e. $\sum_j (\sum_i(<x>_j - x_{ij})^2)$ with $x_{ij}$ an observable of star $j$ determined with method $i$,  $<x>_j$ indicates the mean value computed over all methods of an observable $x$ of star $j$. So we compute for each star the dispersion between results of different methods for both $\Delta \nu$ and $\nu_{\rm max}$ independently. For each cluster / population of field stars, the results of the method with the smallest total deviations from the mean are used in this study. It turned out that the same method gave the lowest deviation for each variable, and hence the results of $\Delta \nu$ and $\nu_{\rm max}$ did not need to be joined. To capture the statistical fluctuations within the selected methods and between the different methods, we computed the final uncertainty in each parameter for each star as the intrinsic uncertainty of the selected method, to which we added in quadrature the dispersion (variance) in the central values returned by all methods. The latter accounts for the fact that the formal errors returned by individual methods are often smaller than the dispersion between the results obtained by the different methods. This procedure provided us with 46, 42, 4 and 662 stars for NGC~6791, NGC~6819, NGC~6811 and the field stars, using methods SYD, OCT, SYD and COR, respectively. Results obtained in the same way and using the same methods have also been used by \citet{basu2011}.

For some stars the global oscillation parameters $\Delta \nu$ and $\nu_{\rm max}$ determined by different methods were not in agreement. These stars are not taken into account in the analysis performed in this work, but we investigated why it was more difficult to obtain consistent oscillation parameters for these stars. In a number of stars, signatures of oscillations at very low frequencies are visible by eye, but determining reliable parameters is not yet possible with the limited timespan of the data we currently have at our disposal. Furthermore, for some of the fainter stars the noise level due to shot noise, at the frequency of the oscillations makes accurate determinations of $\Delta \nu$ and $\nu_{\rm max}$ difficult. Both these issues can be overcome with longer timeseries, which are currently being acquired by \textit{Kepler}.

\subsection{Oscillations and stellar parameters}
As indicated above, for the cluster red giants we have reliable determinations of $\Delta \nu$ and $\nu_{\rm max}$ for 46, 42 and 4 red-giant stars in the clusters NGC~6791, NGC~6819 and NGC~6811,  respectively. For these stars we also have effective temperatures from multicolour photometry. Following the investigations by  \citet{hekker2009,stello2009a,mosser2010} we have fitted a polynomial of the form: 
\begin{equation}
\Delta \nu \approx a \nu_{\rm max}^b,
\label{numaxdnu}
\end{equation}
to the results of each cluster. The values of $a$ and $b$ for the respective clusters and field stars are listed in Table~\ref{abres}. These results show that the values for $b$ are consistent within the uncertainties of the different samples of stars, while the values for $a$ are significantly different. The top left panel of Fig.~\ref{resclusters} shows the correlation between $\Delta \nu$ and $\nu_{\rm max}$ for each of the clusters. Similar results for the field stars are shown in Fig.~\ref{resfield}.

\begin{table}
\begin{minipage}{\linewidth}
\caption{Values for the parameters $a$ and $b$ (Eq.~\ref{numaxdnu}) for the three clusters and the field stars, as well as the number of stars taken into account in the respective clusters and in the field.}
\label{abres}
\centering
\begin{tabular}{cccc}
\hline\hline
cluster    &   $a$   &   $b$  & $N$ \\
\hline
NGC~6791  &  0.28 $\pm$ 0.01 & 0.75 $\pm$ 0.01 & 46\\
NGC~6819  &  0.24 $\pm$ 0.01 & 0.77 $\pm$ 0.01 & 42\\   
NGC~6811 &   0.20 $\pm$ 0.06 & 0.79 $\pm$ 0.06 & 4\\
field & 0.266 $\pm$ 0.004 & 0.761 $\pm$ 0.004 & 662\\
\hline
\end{tabular}
\end{minipage}
\end{table}

With the values of $\Delta \nu$, $\nu_{\rm max}$ and the effective temperatures (see Sect.~2.3) the scaling relations described by \citet{kjeldsen1995} have been used to compute the masses and radii of the stars:
\begin{equation}
\nu_{\rm max} \approx \nu_{\rm max \odot}\frac{M/M_{\odot}}{(R/R_{\odot})^2 \sqrt{T_{\rm eff}/ \rm T_{\rm eff \odot}}},
\label{numax}
\end{equation}
\begin{equation}
\Delta\nu \approx \Delta \nu_{\odot}\sqrt{\frac{M/M_{\odot}}{(R/R_{\odot})^3}},
\label{dnu}
\end{equation}
with T$_{\rm eff \odot}$~=~5777~K, and the solar values $\Delta \nu_{\odot}$~=~134.88~$\mu$Hz and $\nu_{\rm max \odot}$~=~3120~$\mu$Hz taken from \citet{kallinger2010}.

Normalised histograms of both masses and radii are shown in the bottom panels of Fig.~\ref{resclusters}. The radii and temperatures are used to compute the luminosities as $L$~$\propto$~$R^2 T_{\rm eff}^4$, which are used for the H-R diagram shown in the top right panel of Fig.~\ref{resclusters}. Isochrones are also shown for the clusters NGC~6791 and NGC~6819, which are further discussed by \citet{stello2011} and Miglio et al. (in preparation). Results for field stars are shown in gray in the H-R diagram and in the histograms of the mass and radius. These results are obtained from \textit{Kepler} data observed during Q1, Q2, Q3 and Q4 in the same way as the results for the clusters and are similar to results presented previously for red giants observed with \textit{Kepler} \citep{bedding2010,hekker2010,huber2010,kallinger2010} and CoRoT \citep{deridder2009,hekker2009,kallinger2010a,mosser2010}.

\subsection{Influence of reddening on results}
We also investigated the influence of uncertainties in reddening on the computed luminosities, masses and radii. We computed the temperatures of NGC~6819 with $E(B-V)$ of 0.10, 0.15 and 0.18. Using the resulting different temperatures we computed luminosities, masses and radii. We find that for a difference in $E(B-V)$ of 0.01 we get a difference of 23 K, 1.2 L$_{\odot}$, 0.013 M$_{\odot}$ and 0.025 R$_{\odot}$, in $T_{\rm eff}$, luminosity, mass and radius, respectively. Indeed the temperature only appears to the power $1/2$ in Eq.~\ref{numax}, and thus both mass and radius depend only weakly on $T_{\rm eff}$ and the uncertainties on the masses and radii due to reddening are therefore negligible for the clusters, considering reddening uncertainties of the order of 0.01 or less. The luminosity depends strongly on the effective temperature and hence uncertainties due to reddening are non-negligible.

For field stars the reddening is not as well determined as for cluster stars.  Therefore the uncertainties due to reddening on both the temperature and luminosities are not negligible. The rather unstructured shape of the gray dots in the top right panel of Fig.~\ref{resclusters} is most likely caused by large uncertainties in the effective temperatures. With reddening uncertainties of the order of 0.1 or higher, the influence on the masses and radii are most likely also non-negligible. If we assume that $E(B-V)$ values have large uncertainies, but that these are not systematically biased, then the distributions of mass and radius are still valid for field stars, but maybe narrower than shown.

\section{Discussion}
\begin{figure}
\begin{minipage}{\linewidth}
\centering
\includegraphics[width=\linewidth]{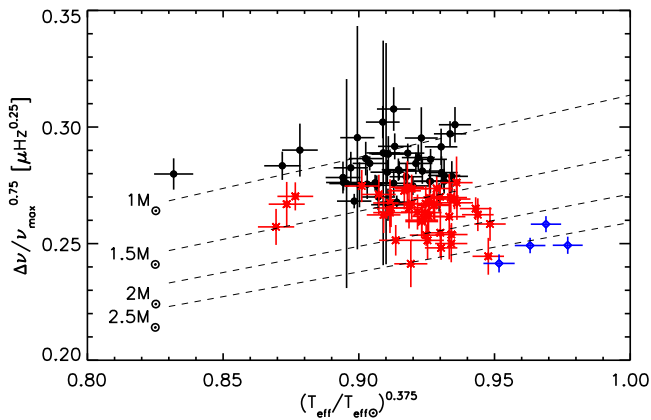}
\end{minipage}
\hfill
\caption{$\Delta \nu / \nu_{\rm max}^{0.75}$  vs. $(T_{\rm eff} / \rm T_{\rm eff \odot})^{0.375}$ for the cluster stars (same colour coding as in Fig.~\ref{resclusters}). The dashed lines indicate isomass lines derived from  Eq.~\ref{ratio} and the observations of the clusters (see text for more details).}
\label{massdiagram}
\end{figure}

\subsection{$\Delta \nu$ vs. $\nu_{\rm max}$ relation}
As already described by \citet{stello2009a} and \citet{huber2010},  the ratio of $\Delta \nu$ to $\nu_{\rm max}^{0.75}$ depends on mass and effective temperature only in the following way:
\begin{equation}
\frac{\Delta \nu}{\nu_{\rm max}^{0.75}} =  \frac{\Delta \nu_{\odot}}{\nu_{\rm max \odot}^{0.75}}\left(\frac{M}{\rm M_{\odot}}\right)^{-0.25} \left(\frac{T_{\rm eff}}{\rm T_{\rm eff \odot}}\right)^{0.375}.
\label{ratio}
\end{equation}
Note that this relation assumes no dependence on metallicity in the scaling relations. Additionally, no adiabatic effects are included in the scaling relations used here, which could explain the slightly higher values of $b$ found in the observations, compared to the value of 0.75 in Eq.~\ref{ratio}.
To investigate this mass dependence further, we first use the field red giants and Eq.~\ref{numaxdnu}. The parameters [$a$,$b$] for the complete sample are roughly mid-way between the values for NGC~6791 and NGC~6819.  We now divide the field sample into a low-mass sample, in which stars with $M$~$<$~1.5~M$_{\odot}$ are selected, and a high-mass sample with $M$~$\geq$~1.5~M$_{\odot}$. The parameters [$a$,$b$] now become [0.282 $\pm$ 0.003,0.754 $\pm$ 0.002], [0.262 $\pm$ 0.005,0.753 $\pm$ 0.005], respectively. These values of $a$ are more in line with those obtained for the lower mass cluster stars in NGC~6791 and the higher mass cluster stars in NGC~6819, respectively. The difference in the value $a$ between the high mass field stars and NGC~6819 could be due to 1)  stars with masses $<$ 1.5~M$_{\odot}$ in NGC~6819 and 2) the different mass distributions.

The mass dependence is also apparent when $\Delta \nu / \nu_{\rm max}^{0.75}$ is plotted as a function of $(T_{\rm eff}/ \rm T_{\rm eff \odot})^{0.375}$ (see Fig.~\ref{massdiagram}). Indeed, the stars of the three clusters occupy different locations in this diagram. Isomass lines are computed from Eq.~\ref{ratio}. This way of presenting the data shows a clear separation of the clusters and allows one to constrain the mass ranges within them. Differences in mass between the clump and red-giant branch stars are investigated by Miglio et al. (in preparation).

\begin{table*}
\begin{minipage}{\linewidth}
\caption{Observed mass and $\nu_{\rm max}$ ranges for the clusters NGC~6791 and NGC~6819 and the parameters $a$ and $b$ derived from observations and models}
\label{simint}
\centering
\begin{tabular}{ccccccc}
\hline\hline
cluster    &   mass   &   $\nu_{\rm max}$  & $a_{\rm obs}$ & $b_{\rm obs}$ & $a_{\rm mod}$ & $b_{\rm mod}$\\
\hline
NGC~6791  &     0.9~$<$~$M/\rm M_{\odot}$~$<$~1.5 & 5~$\mu$Hz~$<$~$\nu_{\rm max}$~$<$~100~$\mu$Hz & 0.28 $\pm$ 0.01 & 0.75 $\pm$ 0.01 & 0.280 $\pm$ 0.001 & 0.753 $\pm$ 0.001 \\
NGC~6819  &     1.4~$<$~$M/\rm M_{\odot}$~$<$~2.1 &  5~$\mu$Hz~$<$~$\nu_{\rm max}$~$<$~150~$\mu$Hz & 0.24 $\pm$ 0.01 & 0.77 $\pm$ 0.01 & 0.256 $\pm$ 0.001 & 0.756 $\pm$ 0.001\\
\hline
\end{tabular}
\end{minipage}
\end{table*}

\begin{figure*}
\begin{minipage}{8.4 cm}
\centering
\includegraphics[width=8.4cm]{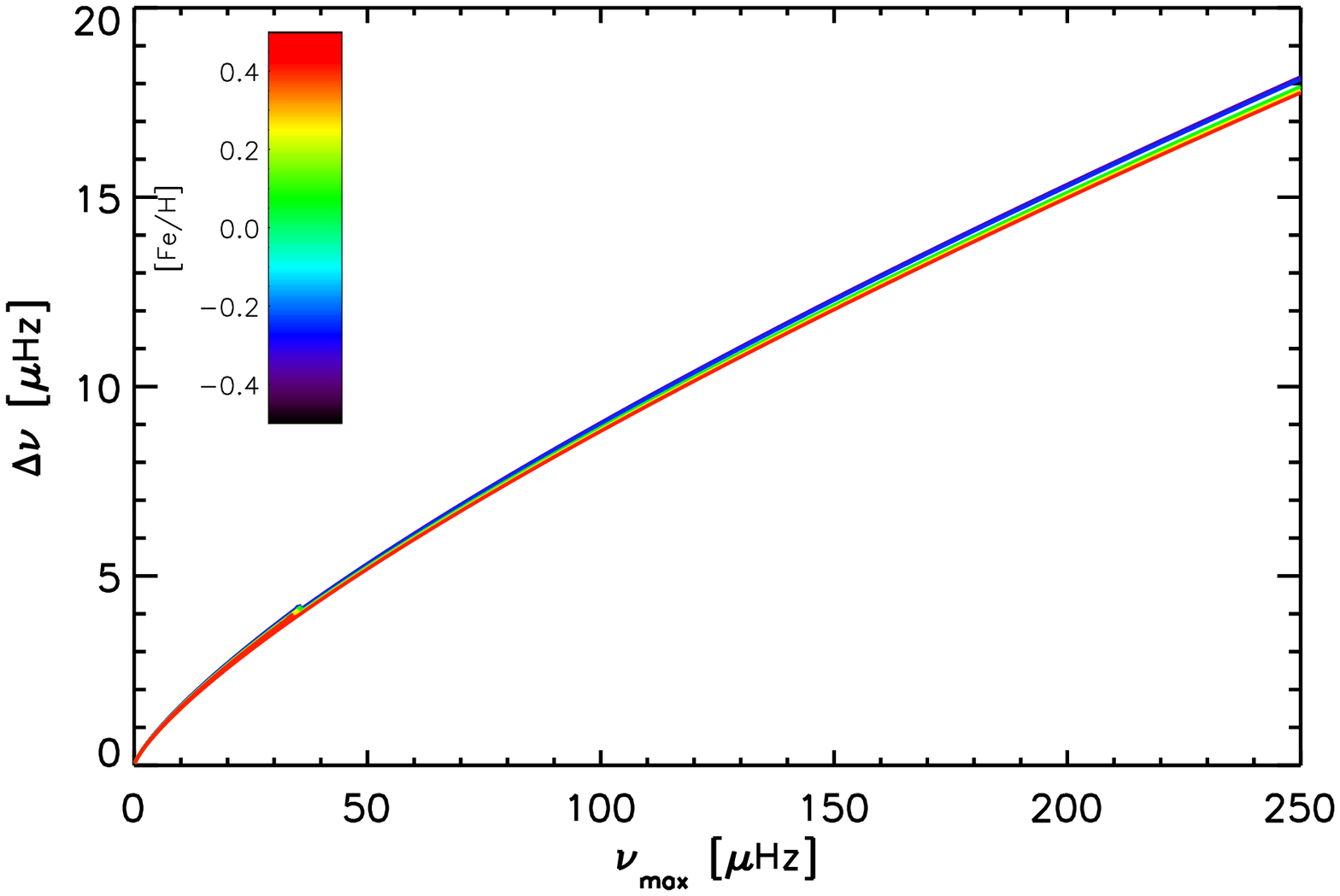}
\end{minipage}
\hfill
\begin{minipage}{8.4 cm}
\centering
\includegraphics[width=8.4cm]{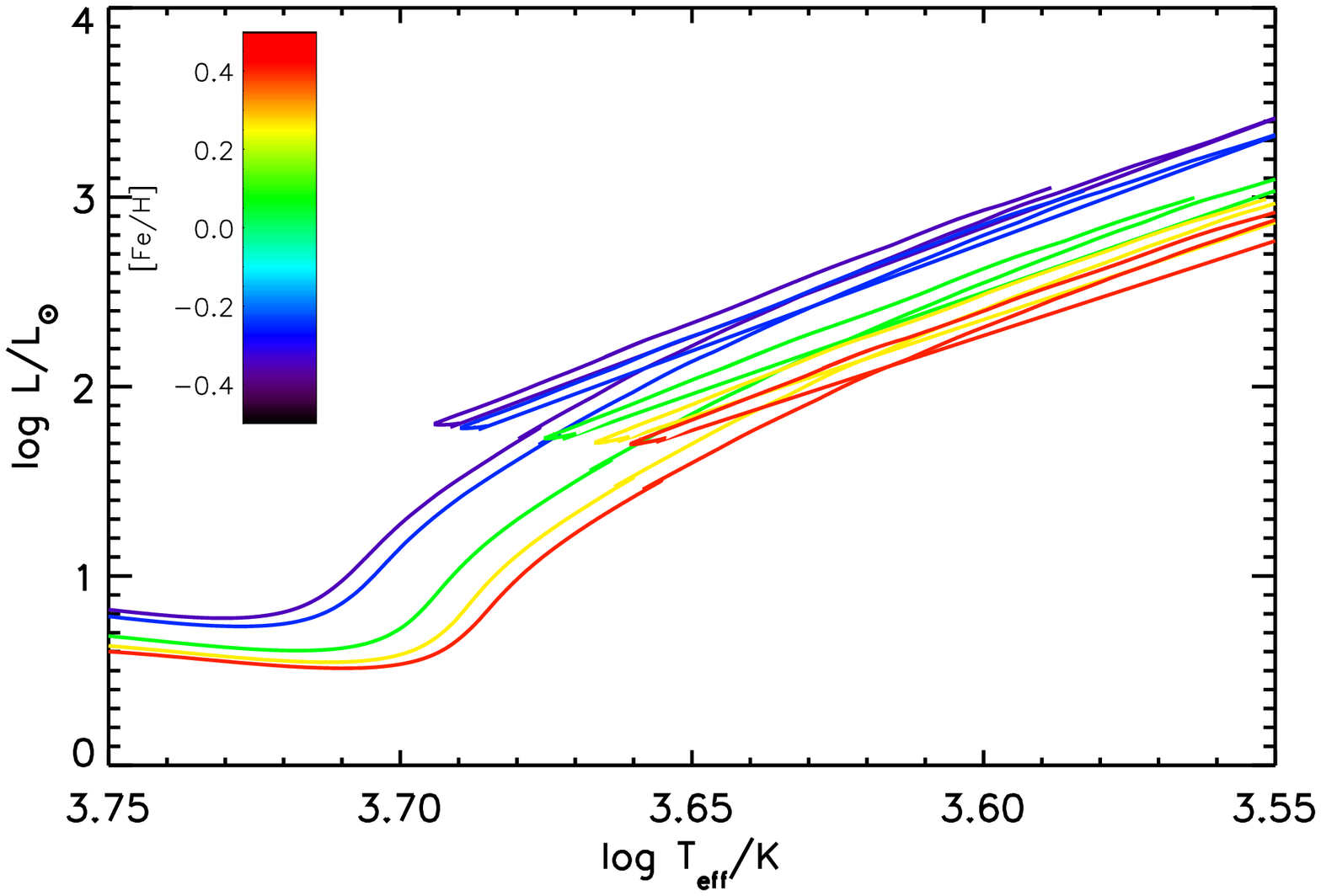}
\end{minipage}
\hfill
\begin{minipage}{8.4cm}
\centering
\includegraphics[width=8.4cm]{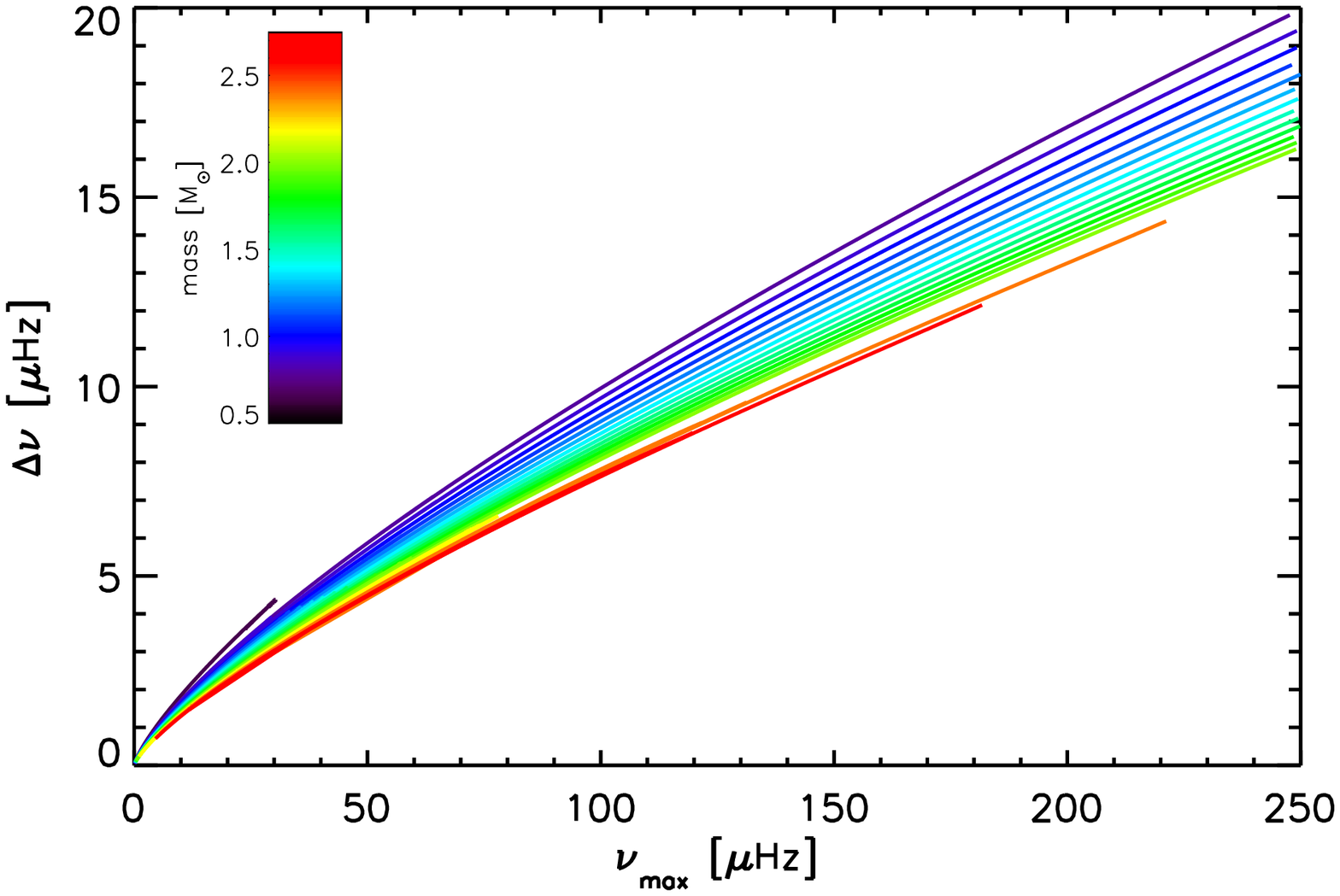}
\end{minipage}
\hfill
\begin{minipage}{8.4cm}
\centering
\includegraphics[width=8.4cm]{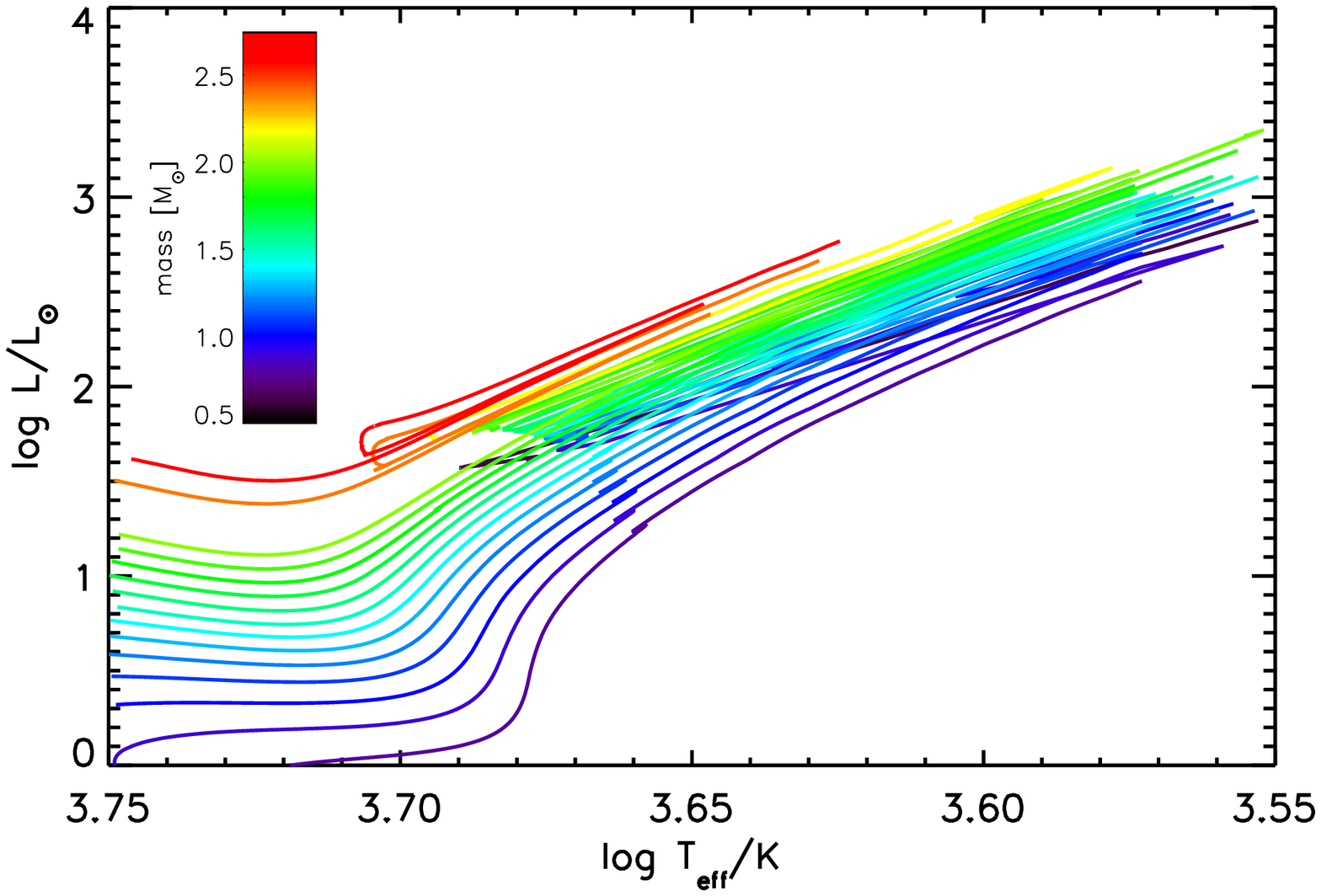}
\end{minipage}
\caption{$\Delta \nu$ vs. $\nu_{\rm max}$ correlations (left) and H-R diagrams (right) obtained from \textit{BaSTI} evolution models with 1.3 M$_{\odot}$ and variable metallicity (top) and with [Fe/H]~=~0.06~dex and variable mass (bottom). Note that in for the $\Delta \nu$ vs. $\nu_{\rm max}$ correlations only stars with $T_{\rm eff}$~$<$~5250~K have been taken into account.}
\label{resmodels}
\end{figure*}

We also investigated the influence of mass and metallicity on the $\Delta \nu$ vs. $\nu_{\rm max}$ relation using \textit{BaSTI} evolution models \citep[e.g.,][]{pietrinferni2004,cordier2007}. We used models with a fixed mass of 1.3~M$_{\odot}$ and a range of metallicities, i.e. [Fe/H]~=~[$+$0.4,$+$0.26,$+$0.06,$-$0.25,$-$0.35,$-$0.66]. The metallicity dependence of the $\Delta \nu - \nu_{\rm max}$ relation is not significant (top left panel of Fig.~\ref{resmodels}). Note that this is true under the assumption that the metallicity does not influence excitation / damping of the oscillations \citep[see e.g.][for investigations of metallicity on excitation and damping]{houdek1999,samadi2010}. Additionally, we selected models with [Fe/H] = 0.06 and masses ranging from 0.5 to 2.7 M$_{\odot}$. The mass dependence of the $\Delta \nu$ - $\nu_{\rm max}$ relation derived from modelling agrees with the scaling relation given in Eq.~\ref{ratio}, i.e. higher masses give lower $\Delta \nu$ values (bottom left panel of Fig.~\ref{resmodels}). Fits through models with mass and $\nu_{\rm max}$ ranges similar to the ones for the stars observed in the respective clusters give for NGC~6791 the same values for $a$ and $b$ in Eq.~\ref{numaxdnu} as the observations (see Table~\ref{simint}), while the values for NGC~6819 are consistent within 2$\sigma$. The range in $\nu_{\rm max}$ taken into account in the fit described in Eq.~\ref{numaxdnu} is important for the determination of the parameters $a$ and $b$ due to the increased effects of mass on stars with higher $\nu_{\rm max}$, i.e. in general less evolved stars. 

The masses of the stars in NGC~6811 are even higher than for NGC~6819 and their position in the top left panel of Fig.~\ref{resclusters} and Fig.~\ref{massdiagram} are consistent with this. The small number of stars, resulting in relatively large uncertainties in $a$ and $b$, is insufficient for a quantitative comparison with fits through the models.

\subsection{Mass and radius distribution}
The mass distributions of the three clusters clearly show that the oldest cluster NGC~6791 contains low-mass stars, whereas the younger clusters NGC~6819 and NGC~6811 contain stars with higher masses. This is as expected because for the younger clusters only the higher mass stars have had time to evolve to giants. For the older cluster the lower-mass stars have also had time to evolve to become giants, while the higher-mass stars have already evolved further and are no longer red giants. We note that the distribution of masses derived in this work for stars in the clusters NGC~6791 and NGC~6819 are consistent with the ones derived using a grid-based method \citep{basu2011}, when taking into account the higher uncertainties in the direct method used here. Additionally, the stellar masses of NGC~6791 are also in agreement with the turn-off mass obtained from the primary component in the binary V20, i.e. $M_{\rm V20}$~=~1.087~$\pm$~0.004~M$_{\odot}$ \citep{brogaard2011}.

The distribution of field giants is peaked at masses of 1.2 to 1.3 M$_{\odot}$. Hardly any field giants with masses lower than the lowest masses of NGC~6791 or as high as the highest mass stars in NGC~6811 are present. This distribution allows us to make a comment on the likely ages of the field stars with respect to the clusters. The low fraction of high-mass giants might be due to the fact that these stars evolve relatively quickly and do not reside very long in a giant phase and thus that the field stars in the \textit{Kepler} field of view are all older than NGC~6811. Similarly, the low fraction of field giants with lower masses than stars in NGC~6791 implies that the oldest observed field stars are of the same age or younger than NGC~6791. Metallicity might also have some influence on the mass distribution through its influence on the effective temperature (see top right plot of Fig.~\ref{resmodels}), which is used to compute stellar masses (Eq.~\ref{numax}). However, the mass only depends on the square-root of the temperature, and this effect is not expected to be significant enough to alter the inference on the relative age of the field stars. Additionally, we expect that the range of metallicities present among the field stars includes the metallicities of the clusters.

The radius distributions of the three clusters and the field stars overlap, with a bimodal distribution most prominent for stars in NGC~6791. In general, the stars with radii in the range 5-9 R$_{\odot}$ are most likely less-evolved H-shell burning stars ascending the red-giant branch, while the stars with radii $\sim$11R$_{\odot}$ are most likely He-core burning red-clump stars \citep{miglio2009,kallinger2010,mosser2010}. This shows that for the clusters, a significant fraction of stars are still in the (less-evolved) H-shell burning phase, while for the field stars the majority of the stars are in the He-core burning red-clump phase. This is also confirmed by the locations of the stars in the colour-magnitude diagram in Fig.~\ref{CMD6791}.

\subsection{H-R diagram}
\begin{figure}
\begin{minipage}{\linewidth}
\centering
\includegraphics[width=\linewidth]{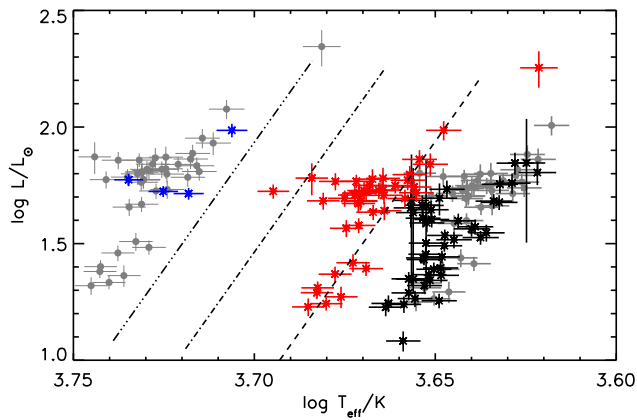}
\end{minipage}
\caption{H-R diagram of the clusters with NGC~6791 in black, NGC~6819 in red and NGC~6811 in blue. The gray symbols indicate NGC~6819 shifted to the positions of NGC~6791 and NGC~6811, respectively. The dashed line roughly indicates the red-giant branch of NGC~6819. The dashed-dotted  line indicates the position of the red-giant branch of NGC~6819 when only the mass is increased to 2.6 M$_{\odot}$ (NGC~6811). The dashed-dotted-dotted-dotted line illustrates the position of the red-giant branch of NGC~6819 when a mass of 2.6 M$_{\odot}$ and a metallicity of $-$0.35~dex are assumed (see text for more details).}
\label{HRclusters}
\end{figure}

The H-R diagrams of the different clusters look very similar but with an offset with respect to one another, most notably in $\log T_{\rm eff}$, but also in $\log L/ \rm L_{\odot}$ (see Fig.~\ref{HRclusters}). For an explanation of this we have again used models. From the right panels of Fig.~\ref{resmodels} it is clear that both mass and metallicity influence the location of a star in the H-R diagram. When leaving all other parameters the same, stars with higher metallicities shift to lower effective temperatures and luminosities, while higher masses give higher effective temperatures and luminosities. NGC~6791 has a significantly higher metallicity and consists of lower mass stars compared to NGC~6819 (see Table~\ref{cluster_param} and lower left panel of Fig.~\ref{resclusters}). So both mass and metallicity add to the separation of the location of the two clusters in the H-R diagram.

To quantify this shift further we computed the change in both effective temperature and luminosity for models in the range 1.0~$<$~$\log L/ \rm L_{\odot}$~$<$~2.0 and 3.6~$<$~$\log T_{\rm eff}$~$<$~3.75 due to a change in mass or metallicity. We change the mass from $\sim$1.7~M$_{\odot}$ (NGC~6819) to $\sim$1.3~M$_{\odot}$ (NGC~6791) for models with constant metallicity of 0.06~dex (similar to the metallicity of NGC~6819). This change in mass induces a change in $\log T_{\rm eff}$ of $-$0.016 and in $\log L/ \rm L_{\odot}$ of 0.05. For the metallicity we compute the difference in $\log T_{\rm eff}$ for models with $M$~=~1.3~M$_{\odot}$ (NGC~6791) due to a metallicity change from 0.06~dex (NGC~6819) to 0.4~dex (NGC~6791). This change in metallicity induces a difference in $\log T_{\rm eff}$ of $-$0.014 and in $\log L/ \rm L_{\odot}$ of 0.03. Applying the total shifts $\log T_{\rm eff}$ = $-$0.03 and $\log L/ \rm L_{\odot}$ = 0.02 to the data of NGC~6819 indeed places the data roughly at the position of the observations of NGC~6791 (see gray dots on the right-hand side of Fig.~\ref{HRclusters}). From this analysis we can conclude that both the metallicity and mass difference between the clusters contribute to the observed shift in the position in the H-R diagram.


For NGC~6811 no direct metallicity determination is available and so far solar metallicity has been assumed, similar to the metallicity of NGC~6819. In that respect the offset of the NGC~6811 compared to NGC~6819 should be mostly due to the difference in mass. From the models with [Fe/H]~=~0.06~dex we find that an increase in mass from 1.7~M$_{\odot}$ (NGC~6819) to 2.6~M$_{\odot}$ (NGC~6811) would cause an offset in $\log T_{\rm eff}$ of the order of 0.03 and an offset in $\log L/ \rm L_{\odot}$ of about 0.05. Shifting the position of the red-giant branch of NGC~6819 (dashed line in Fig.~\ref{HRclusters}) by these amounts would result in the position indicated with the dashed-dotted line in Fig.~\ref{HRclusters}. This location is not consistent with the observations and indicates that the metallicity of NGC~6811 is subsolar. Therefore, we used a metallicity for NGC~6811 of $-$0.35~dex. The additional shift induced by this metallicity would shift the red-giant branch of NGC~6819 to the position indicated with the dashed-dotted-dotted-dotted line in Fig.~\ref{HRclusters}. This could be consistent with the observations, if we assume that the observed stars of NGC~6811 are red-clump stars. However, if the stars in NGC~6811  are ascending the red-giant branch, it would mean that the observed offset of the locations of the stars with respect to NGC~6819 is even larger and a metallicity of $-$0.66~dex is used. This metallicity value would shift the position of NGC~6819 indeed approximately to the position of NGC~6811 (gray dots on the left-hand side of Fig.~\ref{HRclusters}). We thus find that the metallicity of NGC~6811 is subsolar in the range $-$0.3 to $-$0.7 dex. This result is based on observations of only a few stars in the cluster, and a direct determination of the metallicity is needed to confirm which of the described scenarios is to be favoured.\newline

It is also interesting to view the results for the clusters in light of the results obtained by \citet{kallinger2010} from a detailed comparison between field giants and models. They found that most of the red-clump stars have an initial composition similar to the Sun, while stars in the bump are more consistent with metal-enhanced stars. The bump is a higher concentration of stars on the ascending red-giant branch in the H-R diagram \citep{alves1999}. The bump luminosity corresponds to the evolutionary phase during which the outward moving H-burning shell passes through the mass marking the maximum extent of the bottom of the convection zone in slightly earlier evolution phases. At this point there is a discontinuity in the hydrogen abundance, and when the burning shell moves into the region of higher hydrogen abundance the luminosity decreases slightly before the star continues to ascend the giant branch.  The metallicity effect mentioned  by \citet{kallinger2010}  could be caused by a too high mixing-length parameter used to construct the models, i.e., a slightly less efficient convection would shift the solar-metallicity bump models towards lower temperatures in the direction of the observed bump. However, if the metallicity effect is real, it would be very interesting. In addition, \citet{nataf2010} discuss the detectability of the bump in the galactic bulge and compare the bump features with bumps detected in globular clusters. They find that the relative brightness of clump and bump stars is strongly correlated with metallicity. 

One could conclude from this that it would seem that a bump is more likely to be detected in a metal-rich open cluster, such as NGC~6791, than in a solar-metallicity open cluster, such as NGC~6819. However, there are many factors that make it difficult to draw any firm conclusions in this instance.
The sample of red-giant stars in the open clusters investigated here is not very rich; conclusions for field stars and globular clusters are not necessarily correct for stars in open cluster; the findings by \citet{kallinger2010} have yet to be confirmed with additional modelling; and so far no bump has been detected in any open cluster. We should also add that we did not detect a bump in any of the open clusters in the \textit{Kepler} field of view from the current analysis.

\section{Summary and conclusions}
From the global asteroseismic parameters and derived stellar parameters (luminosity, effective temperature, mass, and radius) of red giants in three open clusters NGC~6791, NGC~6819 and NGC~6811 and in the field, we have investigated the influence of evolution and metallicity on the red-giant star populations. From this investigation we conclude the following:
\begin{itemize}
\item Mass has a significant influence on the $\Delta \nu$ - $\nu_{\rm max}$ relation, while the influence of metallicity is negligible, under the assumption that the metallicity does not influence the excitation / damping of the oscillations. This has been predicted from models, but now also clearly shown in observed data.
\item It is well known that both mass and metallicity have influence on the position of stars in the H-R diagram. The different positions of the old metal-rich cluster NGC~6791 and the middle-aged solar-metallicity cluster NGC~6819 can indeed be explained by the observed differences in metallicity and mass. With this confirmation of the theory, we also investigated the metallicity of NGC~6811, for which cluster no direct metallicity measurements are available. The location of the young cluster NGC~6811 cannot be explained if we assume solar metallicity for this cluster. A metallicity of about $-$0.35 dex is needed to explain the position if the observed stars are He-core burning red-clump stars. However, if these stars are H-shell burning stars ascending the red-giant branch, the location of the stars in the H-R diagram can only be explained when the cluster has a subsolar metallicity of about $-$0.7 dex.
\item The distributions of the masses of the cluster stars also allowed us to say something about the relative age of the field stars: the observed field stars are all of the same age or younger than NGC~6791 and older than NGC~6811.
\end{itemize}

\acknowledgements
The authors gratefully acknowledge the \textit{Kepler} Science Team and all those who have contributed to making the \textit{Kepler} mission possible. Funding for the \textit{Kepler} Discovery mission is provided by NASAs Science Mission Directorate. This work has made use of \textit{BaSTI} web tools. SH, WJC, YE and SJH acknowledge financial support from the UK Science and Technology Facilities Council (STFC). SH also acknowledges financial support from the Netherlands Organisation for Scientific Research (NWO).
This project has been supported by the `Lend\"ulet' program of the Hungarian Academy of Sciences and the Hungarian OTKA grants K83790 and MB08C 81013. NCAR is supported by the National Science Foundation.

\bibliographystyle{aa}
\bibliography{16303bib.bib}

\begin{thebibliography}{60}
\expandafter\ifx\csname natexlab\endcsname\relax\def\natexlab#1{#1}\fi

\bibitem[{{Alves} \& {Sarajedini}(1999)}]{alves1999}
{Alves}, D.~R. \& {Sarajedini}, A. 1999, \apj, 511, 225

\bibitem[{{Anthony-Twarog} {et~al.}(2007){Anthony-Twarog}, {Twarog}, \&
  {Mayer}}]{anthony2007}
{Anthony-Twarog}, B.~J., {Twarog}, B.~A., \& {Mayer}, L. 2007, \aj, 133, 1585

\bibitem[{{Baglin} {et~al.}(2006){Baglin}, {Auvergne}, {Barge}, {Deleuil},
  {Catala}, {Michel}, {Weiss}, \& {The COROT Team}}]{baglin2006}
{Baglin}, A., {Auvergne}, M., {Barge}, P., {et~al.} 2006, in ESA Special
  Publication, Vol. 1306, ESA Special Publication, ed. M.~{Fridlund},
  A.~{Baglin}, J.~{Lochard}, \& L.~{Conroy}, 33

\bibitem[{{Basu} {et~al.}(2011){Basu}, {Grundahl}, {Stello}, {Kallinger},
  {Hekker}, {Mosser}, {Garc{\'{\i}}a}, {Mathur}, {Brogaard}, {Bruntt},
  {Chaplin}, {Gai}, {Elsworth}, {Esch}, {Ballot}, {Bedding}, {Gruberbauer},
  {Huber}, {Miglio}, {Yildiz}, {Kjeldsen}, {Christensen-Dalsgaard},
  {Gilliland}, {Fanelli}, {Ibrahim}, \& {Smith}}]{basu2011}
{Basu}, S., {Grundahl}, F., {Stello}, D., {et~al.} 2011, \apjl, 729, L10

\bibitem[{{Bedding} {et~al.}(2010){Bedding}, {Huber}, {Stello}, {Elsworth},
  {Hekker}, {Kallinger}, {Mathur}, {Mosser}, {Preston}, {Ballot}, {Barban},
  {Broomhall}, {Buzasi}, {Chaplin}, {Garc{\'{\i}}a}, {Gruberbauer}, {Hale}, {De
  Ridder}, {Frandsen}, {Borucki}, {Brown}, {Christensen-Dalsgaard},
  {Gilliland}, {Jenkins}, {Kjeldsen}, {Koch}, {Belkacem}, {Bildsten}, {Bruntt},
  {Campante}, {Deheuvels}, {Derekas}, {Dupret}, {Goupil}, {Hatzes}, {Houdek},
  {Ireland}, {Jiang}, {Karoff}, {Kiss}, {Lebreton}, {Miglio}, {Montalb{\'a}n},
  {Noels}, {Roxburgh}, {Sangaralingam}, {Stevens}, {Suran}, {Tarrant}, \&
  {Weiss}}]{bedding2010}
{Bedding}, T.~R., {Huber}, D., {Stello}, D., {et~al.} 2010, \apjl, 713, L176

\bibitem[{{Bedin} {et~al.}(2008){Bedin}, {King}, {Anderson}, {Piotto},
  {Salaris}, {Cassisi}, \& {Serenelli}}]{bedin2008}
{Bedin}, L.~R., {King}, I.~R., {Anderson}, J., {et~al.} 2008, \apj, 678, 1279

\bibitem[{{Bedin} {et~al.}(2005){Bedin}, {Salaris}, {Piotto}, {King},
  {Anderson}, {Cassisi}, \& {Momany}}]{bedin2005}
{Bedin}, L.~R., {Salaris}, M., {Piotto}, G., {et~al.} 2005, \apjl, 624, L45

\bibitem[{{Borucki} {et~al.}(2009){Borucki}, {Koch}, {Batalha}, {Caldwell},
  {Christensen-Dalsgaard}, {Cochran}, {Dunham}, {Gautier}, {Geary},
  {Gilliland}, {Jenkins}, {Kjeldsen}, {Lissauer}, \& {Rowe}}]{borucki2009}
{Borucki}, W., {Koch}, D., {Batalha}, N., {et~al.} 2009, in IAU Symposium, Vol.
  253, IAU Symposium, 289--299

\bibitem[{{Bragaglia} {et~al.}(2001){Bragaglia}, {Carretta}, {Gratton}, {Tosi},
  {Bonanno}, {Bruno}, {Cal{\`i}}, {Claudi}, {Cosentino}, {Desidera},
  {Farisato}, {Rebeschini}, \& {Scuderi}}]{bragaglia2001}
{Bragaglia}, A., {Carretta}, E., {Gratton}, R.~G., {et~al.} 2001, \aj, 121, 327

\bibitem[{{Brogaard} {et~al.}(2011){Brogaard}, {Bruntt}, {Grundahl}, {Clausen},
  {Frandsen}, {Vandenberg}, \& {Bedin}}]{brogaard2011}
{Brogaard}, K., {Bruntt}, H., {Grundahl}, F., {et~al.} 2011, \aap, 525, A2

\bibitem[{{Brown} {et~al.}(2011){Brown}, {Latham}, {Everett}, \&
  {Esquerdo}}]{brown2011}
{Brown}, T.~M., {Latham}, D.~W., {Everett}, M.~E., \& {Esquerdo}, G.~A. 2011,
  ArXiv e-prints:1102.0342

\bibitem[{{Carraro} {et~al.}(2006){Carraro}, {Villanova}, {Demarque},
  {McSwain}, {Piotto}, \& {Bedin}}]{carraro2006}
{Carraro}, G., {Villanova}, S., {Demarque}, P., {et~al.} 2006, \apj, 643, 1151

\bibitem[{{Carretta} {et~al.}(2007){Carretta}, {Bragaglia}, \&
  {Gratton}}]{carretta2007}
{Carretta}, E., {Bragaglia}, A., \& {Gratton}, R.~G. 2007, \aap, 473, 129

\bibitem[{{Cordier} {et~al.}(2007){Cordier}, {Pietrinferni}, {Cassisi}, \&
  {Salaris}}]{cordier2007}
{Cordier}, D., {Pietrinferni}, A., {Cassisi}, S., \& {Salaris}, M. 2007, \aj,
  133, 468

\bibitem[{{De Ridder} {et~al.}(2009){De Ridder}, {Barban}, {Baudin}, {Carrier},
  {Hatzes}, {Hekker}, {Kallinger}, {Weiss}, {Baglin}, {Auvergne}, {Samadi},
  {Barge}, \& {Deleuil}}]{deridder2009}
{De Ridder}, J., {Barban}, C., {Baudin}, F., {et~al.} 2009, \nat, 459, 398

\bibitem[{{Durgapal} \& {Pandey}(2001)}]{durgapal2001}
{Durgapal}, A.~K. \& {Pandey}, A.~K. 2001, \aap, 375, 840

\bibitem[{{Edmonds} \& {Gilliland}(1996)}]{edmonds1996}
{Edmonds}, P.~D. \& {Gilliland}, R.~L. 1996, \apjl, 464, L157

\bibitem[{{Frolov} {et~al.}(2010){Frolov}, {Ananjevskaja}, {Gorshanov}, \&
  {Polyakov}}]{frolov2010}
{Frolov}, V.~N., {Ananjevskaja}, Y.~K., {Gorshanov}, D.~L., \& {Polyakov},
  E.~V. 2010, Astronomy Letters, 36, 338

\bibitem[{{Garc{\'{\i}}a} {et~al.}(2011){Garc{\'{\i}}a}, {Hekker}, {Stello},
  {Guti\'errez-Soto}, {Handberg}, {Huber}, {Karoff}, {Uytterhoeven},
  {Appourchaux}, {Chaplin}, {Elsworth}, {Mathur}, \& {Ballot}}]{garcia2011}
{Garc{\'{\i}}a}, R.~A., {Hekker}, S., {Stello}, D., {et~al.} 2011, MNRAS,
  submitted

\bibitem[{{Garc{\'{\i}}a} {et~al.}(2005){Garc{\'{\i}}a}, {Turck-Chi{\`e}ze},
  {Boumier}, {Robillot}, {Bertello}, {Charra}, {Dzitko}, {Gabriel},
  {Jim{\'e}nez-Reyes}, {Pall{\'e}}, {Renaud}, {Roca Cort{\'e}s}, \&
  {Ulrich}}]{garcia2005}
{Garc{\'{\i}}a}, R.~A., {Turck-Chi{\`e}ze}, S., {Boumier}, P., {et~al.} 2005,
  \aap, 442, 385

\bibitem[{{Glushkova} {et~al.}(1999){Glushkova}, {Batyrshinova}, \&
  {Ibragimov}}]{glushkova1999}
{Glushkova}, E.~V., {Batyrshinova}, V.~M., \& {Ibragimov}, M.~A. 1999,
  Astronomy Letters, 25, 86

\bibitem[{{Grundahl} {et~al.}(2008){Grundahl}, {Clausen}, {Hardis}, \&
  {Frandsen}}]{grundahl2008}
{Grundahl}, F., {Clausen}, J.~V., {Hardis}, S., \& {Frandsen}, S. 2008, \aap,
  492, 171

\bibitem[{{Hekker} {et~al.}(2010){Hekker}, {Broomhall}, {Chaplin}, {Elsworth},
  {Fletcher}, {New}, {Arentoft}, {Quirion}, \& {Kjeldsen}}]{hekker2010b}
{Hekker}, S., {Broomhall}, A., {Chaplin}, W.~J., {et~al.} 2010, \mnras, 402,
  2049

\bibitem[{{Hekker} {et~al.}(2011){Hekker}, {Elsworth}, {De Ridder}, {Mosser},
  {Garc{\'{\i}}a}, {Kallinger}, {Mathur}, {Huber}, {Buzasi}, {Preston}, {Hale},
  {Ballot}, {Chaplin}, {R{\'e}gulo}, {Bedding}, {Stello}, {Borucki}, {Koch},
  {Jenkins}, {Allen}, {Gilliland}, {Kjeldsen}, \&
  {Christensen-Dalsgaard}}]{hekker2010}
{Hekker}, S., {Elsworth}, Y., {De Ridder}, J., {et~al.} 2011, \aap, 525, A131

\bibitem[{{Hekker} {et~al.}(2009){Hekker}, {Kallinger}, {Baudin}, {De Ridder},
  {Barban}, {Carrier}, {Hatzes}, {Weiss}, \& {Baglin}}]{hekker2009}
{Hekker}, S., {Kallinger}, T., {Baudin}, F., {et~al.} 2009, \aap, 506, 465

\bibitem[{{Hole} {et~al.}(2009){Hole}, {Geller}, {Mathieu}, {Platais},
  {Meibom}, \& {Latham}}]{hole2009}
{Hole}, K.~T., {Geller}, A.~M., {Mathieu}, R.~D., {et~al.} 2009, \aj, 138, 159

\bibitem[{{Houdek} {et~al.}(1999){Houdek}, {Balmforth},
  {Christensen-Dalsgaard}, \& {Gough}}]{houdek1999}
{Houdek}, G., {Balmforth}, N.~J., {Christensen-Dalsgaard}, J., \& {Gough},
  D.~O. 1999, \aap, 351, 582

\bibitem[{{Huber} {et~al.}(2010){Huber}, {Bedding}, {Stello}, {Mosser},
  {Mathur}, {Kallinger}, {Hekker}, {Elsworth}, {Buzasi}, {De Ridder},
  {Gilliland}, {Kjeldsen}, {Chaplin}, {Garc{\'{\i}}a}, {Hale}, {Preston},
  {White}, {Borucki}, {Christensen-Dalsgaard}, {Clarke}, {Jenkins}, \&
  {Koch}}]{huber2010}
{Huber}, D., {Bedding}, T.~R., {Stello}, D., {et~al.} 2010, \apj, 723, 1607

\bibitem[{{Huber} {et~al.}(2009){Huber}, {Stello}, {Bedding}, {Chaplin},
  {Arentoft}, {Quirion}, \& {Kjeldsen}}]{huber2009}
{Huber}, D., {Stello}, D., {Bedding}, T.~R., {et~al.} 2009, Communications in
  Asteroseismology, 160, 74

\bibitem[{{Jenkins} {et~al.}(2010{\natexlab{a}}){Jenkins}, {Caldwell},
  {Chandrasekaran}, {Twicken}, {Bryson}, {Quintana}, {Clarke}, {Li}, {Allen},
  {Tenenbaum}, {Wu}, {Klaus}, {Middour}, {Cote}, {McCauliff}, {Girouard},
  {Gunter}, {Wohler}, {Sommers}, {Hall}, {Uddin}, {Wu}, {Bhavsar}, {Van Cleve},
  {Pletcher}, {Dotson}, {Haas}, {Gilliland}, {Koch}, \&
  {Borucki}}]{jenkins2010}
{Jenkins}, J.~M., {Caldwell}, D.~A., {Chandrasekaran}, H., {et~al.}
  2010{\natexlab{a}}, \apjl, 713, L87

\bibitem[{{Jenkins} {et~al.}(2010{\natexlab{b}}){Jenkins}, {Caldwell},
  {Chandrasekaran}, {Twicken}, {Bryson}, {Quintana}, {Clarke}, {Li}, {Allen},
  {Tenenbaum}, {Wu}, {Klaus}, {Van Cleve}, {Dotson}, {Haas}, {Gilliland},
  {Koch}, \& {Borucki}}]{jenkins2010a}
{Jenkins}, J.~M., {Caldwell}, D.~A., {Chandrasekaran}, H., {et~al.}
  2010{\natexlab{b}}, \apjl, 713, L120

\bibitem[{{Kalirai} {et~al.}(2007){Kalirai}, {Bergeron}, {Hansen}, {Kelson},
  {Reitzel}, {Rich}, \& {Richer}}]{kalirai2007}
{Kalirai}, J.~S., {Bergeron}, P., {Hansen}, B.~M.~S., {et~al.} 2007, \apj, 671,
  748

\bibitem[{{Kalirai} {et~al.}(2008){Kalirai}, {Hansen}, {Kelson}, {Reitzel},
  {Rich}, \& {Richer}}]{kalirai2008}
{Kalirai}, J.~S., {Hansen}, B.~M.~S., {Kelson}, D.~D., {et~al.} 2008, \apj,
  676, 594

\bibitem[{{Kalirai} {et~al.}(2001){Kalirai}, {Richer}, {Fahlman}, {Cuillandre},
  {Ventura}, {D'Antona}, {Bertin}, {Marconi}, \& {Durrell}}]{kalirai2001}
{Kalirai}, J.~S., {Richer}, H.~B., {Fahlman}, G.~G., {et~al.} 2001, \aj, 122,
  266

\bibitem[{{Kalirai} \& {Tosi}(2004)}]{kalirai2004}
{Kalirai}, J.~S. \& {Tosi}, M. 2004, \mnras, 351, 649

\bibitem[{{Kallinger} {et~al.}(2010{\natexlab{a}}){Kallinger}, {Mosser},
  {Hekker}, {Huber}, {Stello}, {Mathur}, {Basu}, {Bedding}, {Chaplin}, {De
  Ridder}, {Elsworth}, {Frandsen}, {Garc{\'{\i}}a}, {Gruberbauer}, {Matthews},
  {Borucki}, {Bruntt}, {Christensen-Dalsgaard}, {Gilliland}, {Kjeldsen}, \&
  {Koch}}]{kallinger2010}
{Kallinger}, T., {Mosser}, B., {Hekker}, S., {et~al.} 2010{\natexlab{a}}, \aap,
  522, A1

\bibitem[{{Kallinger} {et~al.}(2010{\natexlab{b}}){Kallinger}, {Weiss},
  {Barban}, {Baudin}, {Cameron}, {Carrier}, {De Ridder}, {Goupil},
  {Gruberbauer}, {Hatzes}, {Hekker}, {Samadi}, \& {Deleuil}}]{kallinger2010a}
{Kallinger}, T., {Weiss}, W.~W., {Barban}, C., {et~al.} 2010{\natexlab{b}},
  \aap, 509, A77

\bibitem[{{Kang} \& {Ann}(2002)}]{kang2002}
{Kang}, Y. \& {Ann}, H.~B. 2002, Journal of Korean Astronomical Society, 35, 87

\bibitem[{{Kjeldsen} \& {Bedding}(1995)}]{kjeldsen1995}
{Kjeldsen}, H. \& {Bedding}, T.~R. 1995, \aap, 293, 87

\bibitem[{{Landsman} {et~al.}(1998){Landsman}, {Bohlin}, {Neff}, {O'Connell},
  {Roberts}, {Smith}, \& {Stecher}}]{landsman1998}
{Landsman}, W., {Bohlin}, R.~C., {Neff}, S.~G., {et~al.} 1998, \aj, 116, 789

\bibitem[{{Liebert} {et~al.}(1994){Liebert}, {Saffer}, \&
  {Green}}]{liebert1994}
{Liebert}, J., {Saffer}, R.~A., \& {Green}, E.~M. 1994, \aj, 107, 1408

\bibitem[{{Mathur} {et~al.}(2010){Mathur}, {Garc{\'{\i}}a}, {R{\'e}gulo},
  {Creevey}, {Ballot}, {Salabert}, {Arentoft}, {Quirion}, {Chaplin}, \&
  {Kjeldsen}}]{mathur2010}
{Mathur}, S., {Garc{\'{\i}}a}, R.~A., {R{\'e}gulo}, C., {et~al.} 2010, \aap,
  511, A46

\bibitem[{{Meunier} {et~al.}(2007){Meunier}, {Deleuil}, {Moutou}, {Ouchani},
  {Savalle}, \& {Surace}}]{meunier2007}
{Meunier}, J., {Deleuil}, M., {Moutou}, C., {et~al.} 2007, in Astronomical
  Society of the Pacific Conference Series, Vol. 376, Astronomical Data
  Analysis Software and Systems XVI, ed. {R.~A.~Shaw, F.~Hill, \& D.~J.~Bell},
  339

\bibitem[{{Miglio} {et~al.}(2009){Miglio}, {Montalb{\'a}n}, {Baudin},
  {Eggenberger}, {Noels}, {Hekker}, {De Ridder}, {Weiss}, \&
  {Baglin}}]{miglio2009}
{Miglio}, A., {Montalb{\'a}n}, J., {Baudin}, F., {et~al.} 2009, \aap, 503, L21

\bibitem[{{Molenda-{\.Z}akowicz} {et~al.}(2010){Molenda-{\.Z}akowicz},
  {Bruntt}, {Sousa}, {Frasca}, {Biazzo}, {Huber}, {Ireland}, {Bedding},
  {Stello}, {Uytterhoeven}, {Dreizler}, {De Cat}, {Briquet}, {Catanzaro},
  {Karoff}, {Frandsen}, \& {Spezzi}}]{molenda2010}
{Molenda-{\.Z}akowicz}, J., {Bruntt}, H., {Sousa}, S., {et~al.} 2010,
  Astronomische Nachrichten, 331, 981

\bibitem[{{Mosser} \& {Appourchaux}(2009)}]{mosser2009}
{Mosser}, B. \& {Appourchaux}, T. 2009, \aap, 508, 877

\bibitem[{{Mosser} {et~al.}(2010){Mosser}, {Belkacem}, {Goupil}, {Miglio},
  {Morel}, {Barban}, {Baudin}, {Hekker}, {Samadi}, {De Ridder}, {Weiss},
  {Auvergne}, \& {Baglin}}]{mosser2010}
{Mosser}, B., {Belkacem}, K., {Goupil}, M., {et~al.} 2010, \aap, 517, A22

\bibitem[{{Nataf} {et~al.}(2010){Nataf}, {Udalski}, {Gould}, \&
  {Pinsonneault}}]{nataf2010}
{Nataf}, D.~M., {Udalski}, A., {Gould}, A., \& {Pinsonneault}, M.~H. 2010,
  ArXiv e-prints:1011.4293

\bibitem[{{Origlia} {et~al.}(2006){Origlia}, {Valenti}, {Rich}, \&
  {Ferraro}}]{origlia2006}
{Origlia}, L., {Valenti}, E., {Rich}, R.~M., \& {Ferraro}, F.~R. 2006, \apj,
  646, 499

\bibitem[{{Pietrinferni} {et~al.}(2004){Pietrinferni}, {Cassisi}, {Salaris}, \&
  {Castelli}}]{pietrinferni2004}
{Pietrinferni}, A., {Cassisi}, S., {Salaris}, M., \& {Castelli}, F. 2004, \apj,
  612, 168

\bibitem[{{Ram{\'{\i}}rez} \& {Mel{\'e}ndez}(2005)}]{ramirez2005}
{Ram{\'{\i}}rez}, I. \& {Mel{\'e}ndez}, J. 2005, \apj, 626, 465

\bibitem[{{Samadi} {et~al.}(2010){Samadi}, {Ludwig}, {Belkacem}, {Goupil}, \&
  {Dupret}}]{samadi2010}
{Samadi}, R., {Ludwig}, H., {Belkacem}, K., {Goupil}, M.~J., \& {Dupret}, M.
  2010, \aap, 509, A15

\bibitem[{{Skrutskie} {et~al.}(2006){Skrutskie}, {Cutri}, {Stiening},
  {Weinberg}, {Schneider}, {Carpenter}, {Beichman}, {Capps}, {Chester},
  {Elias}, {Huchra}, {Liebert}, {Lonsdale}, {Monet}, {Price}, {Seitzer},
  {Jarrett}, {Kirkpatrick}, {Gizis}, {Howard}, {Evans}, {Fowler}, {Fullmer},
  {Hurt}, {Light}, {Kopan}, {Marsh}, {McCallon}, {Tam}, {Van Dyk}, \&
  {Wheelock}}]{skrutskie2006}
{Skrutskie}, M.~F., {Cutri}, R.~M., {Stiening}, R., {et~al.} 2006, \aj, 131,
  1163

\bibitem[{{Stello} {et~al.}(2010){Stello}, {Basu}, {Bruntt}, {Mosser},
  {Stevens}, {Brown}, {Christensen-Dalsgaard}, {Gilliland}, {Kjeldsen},
  {Arentoft}, {Ballot}, {Barban}, {Bedding}, {Chaplin}, {Elsworth},
  {Garc{\'{\i}}a}, {Goupil}, {Hekker}, {Huber}, {Mathur}, {Meibom},
  {Sangaralingam}, {Baldner}, {Belkacem}, {Biazzo}, {Brogaard}, {Su{\'a}rez},
  {D'Antona}, {Demarque}, {Esch}, {Gai}, {Grundahl}, {Lebreton}, {Jiang},
  {Jevtic}, {Karoff}, {Miglio}, {Molenda-{\.Z}akowicz}, {Montalb{\'a}n},
  {Noels}, {Roca Cort{\'e}s}, {Roxburgh}, {Serenelli}, {Silva Aguirre},
  {Sterken}, {Stine}, {Szab{\'o}}, {Weiss}, {Borucki}, {Koch}, \&
  {Jenkins}}]{stello2010}
{Stello}, D., {Basu}, S., {Bruntt}, H., {et~al.} 2010, \apjl, 713, L182

\bibitem[{{Stello} {et~al.}(2007){Stello}, {Bruntt}, {Kjeldsen}, {Bedding},
  {Arentoft}, {Gilliland}, {Nuspl}, {Kim}, {Kang}, {Koo}, {Lee}, {Sterken},
  {Lee}, {Jensen}, {Jacob}, {Szab{\'o}}, {Frandsen}, {Csubry}, {Dind},
  {Bouzid}, {Dall}, \& {Kiss}}]{stello2007}
{Stello}, D., {Bruntt}, H., {Kjeldsen}, H., {et~al.} 2007, \mnras, 377, 584

\bibitem[{{Stello} {et~al.}(2009){Stello}, {Chaplin}, {Basu}, {Elsworth}, \&
  {Bedding}}]{stello2009a}
{Stello}, D., {Chaplin}, W.~J., {Basu}, S., {Elsworth}, Y., \& {Bedding}, T.~R.
  2009, \mnras, 400, L80

\bibitem[{{Stello} \& {Gilliland}(2009)}]{stello2009}
{Stello}, D. \& {Gilliland}, R.~L. 2009, \apj, 700, 949

\bibitem[{{Stello} {et~al.}(2011){Stello}, {Meibom}, {Gilliland}, {Grundahl},
  {Hekker}, {Mosser}, {Kallinger}, {Mathur}, {Garc\'ia}, {Huber}, {Basu},
  {Bedding}, {Chaplin}, {Elsworth}, {Molenda-Zakowicz}, {Szab\'o}, {Still},
  {Jenkins}, {Christensen-Dalsgaard}, {Kjeldsen}, \& {Serenelli}}]{stello2011}
{Stello}, D., {Meibom}, S., {Gilliland}, R.~L., {et~al.} 2011, \apj, submitted

\bibitem[{{Stetson} {et~al.}(2003){Stetson}, {Bruntt}, \&
  {Grundahl}}]{stetson2003}
{Stetson}, P.~B., {Bruntt}, H., \& {Grundahl}, F. 2003, \pasp, 115, 413

\bibitem[{{Uytterhoeven} {et~al.}(2010){Uytterhoeven}, {Briquet}, {Bruntt}, {De
  Cat}, {Frandsen}, {Guti{\'e}rrez-Soto}, {Kiss}, {Kurtz}, {Marconi},
  {Molenda-{\.Z}akowicz}, {{\O}stensen}, {Randall}, {Southworth}, \&
  {Szab{\'o}}}]{uytterhoeven2010}
{Uytterhoeven}, K., {Briquet}, M., {Bruntt}, H., {et~al.} 2010, Astronomische
  Nachrichten, 331, 993

\end{thebibliography}

\end{document}